\documentclass[prd,preprintnumbers,superscriptaddress,nofootinbib,twocolumn]{revtex4-1}

\usepackage{graphicx}
\usepackage{amsmath}
\usepackage{amssymb}
\usepackage{hyperref}

\allowdisplaybreaks[1] 

\def \refsec#1{Sec.~\ref{#1}}

\def \refapp#1{App.~\ref{#1}}
\def \reffig#1{Fig.~\ref{#1}}
\def \reftab#1{Tab.~\ref{#1}}

\newcommand{\MS}{\overline{\rm MS}}
\newcommand{\OS}{{\text{on-shell}}}

\newcommand{\GF}[1]{G_{\!F}^{#1}}

\begin{document}

\title{\texorpdfstring{%
       \boldmath Electroweak Corrections to $\boldsymbol{B_{s,d}\to \ell^+\ell^-}$}{%
       Electroweak Corrections to B(sd) -> l+ l-}}

\author{Christoph Bobeth} 
       \email[]{christoph.bobeth@ph.tum.de}
	\affiliation{
       			Excellence Cluster Universe, 
       			Technische Universit\"at M\"unchen,
       			D--85748 Garching, Germany}
	\affiliation{
       			Institute for Advanced Study, Lichtenbergstrasse 2a,\\ 
       			Technische Universit\"at M\"unchen, D--85748 Garching, Germany}
\author{Martin Gorbahn} 
       \email[]{martin.gorbahn@liverpool.ac.uk}
	\affiliation{
       			Excellence Cluster Universe,
       			Technische Universit\"at M\"unchen,
       			D--85748 Garching, Germany}
	\affiliation{
       			Department of Mathematical Sciences, 
       			University of Liverpool, 
       			Liverpool L69 3BX, 
       			United Kingdom}
			
\author{Emmanuel Stamou} 
       \email[]{emmanuel.stamou@weizmann.ac.il}
	\affiliation{
       			Excellence Cluster Universe,
       			Technische Universit\"at M\"unchen,
       			D--85748 Garching, Germany}
	\affiliation{
       			Institute for Advanced Study, Lichtenbergstrasse 2a,\\ 
       			Technische Universit\"at M\"unchen, D--85748 Garching, Germany}
        \affiliation{   
                        Department of Particle Physics and Astrophysics,	
			Weizmann Institute of Science,
			Rehovot 76100, Israel}

\preprint{FLAVOUR(267104)-ERC-54}
\preprint{LTH 991}

\begin{abstract}
  We calculate the full two-loop electroweak matching corrections to the
  operator governing the decay $B_q \to \ell^+\ell^-$ in the Standard
  Model. Their inclusion removes an electroweak scheme and scale uncertainty of
  about $\pm 7$\% of the branching ratio. Using different renormalization schemes
  of the involved electroweak parameters, we estimate residual perturbative
  electroweak and QED uncertainties to be less than $\pm 1$\% at the level of
  the branching ratio.
\end{abstract}

\maketitle

%
%
\section{Introduction}

The rare decays of $B_q \to \ell^+\ell^-$ with $q = d,s$ and $\ell = e, \mu,
\tau$ are helicity suppressed in the Standard Model (SM) and can be predicted
with high precision, which turns them into powerful probes of nonstandard
interactions. In November 2012, LHCb \cite{Aaij:2012nna} reported first
experimental evidence of the decay $B_s\to \mu^+\mu^-$ with a signal
significance of $3.5\,\sigma$ and the time integrated and CP-averaged 
branching ratio
\begin{align}
  \label{eq:BR:exp}
  \overline{{\rm Br}}(B_s \to \mu^+\mu^-) & 
  = \left( 3.2^{+1.4}_{-1.2}({\rm stat})^{+0.5}_{-0.3}({\rm sys})\right) \cdot 10^{-9}\,,
\end{align}
well in agreement with SM predictions. More recently, the signal significance
was raised to $4.0\,\sigma$ after analyzing the currently available data
set of $1$ fb$^{-1}$ at $\sqrt{s} = 7$ TeV and $2$ fb$^{-1}$ at $\sqrt{s} = 8$
TeV, with the result \cite{Aaij:2013aka}
\begin{align}
  \label{eq:BR:exp:LHCb2}
  \overline{{\rm Br}}(B_s \to \mu^+\mu^-) & 
  = \left( 2.9^{+1.1}_{-1.0} ({\rm stat})^{+0.3}_{-0.1}({\rm sys}) \right) \cdot 10^{-9}\,.
\end{align}
CMS confirmed this independently utilizing the complete data set of $5$
fb$^{-1}$ at $\sqrt{s} = 7$ TeV and $20$ fb$^{-1}$ at $\sqrt{s} = 8$ TeV
\cite{Chatrchyan:2013bka} obtaining
\begin{align}
  \label{eq:BR:exp:CMS}
  \overline{{\rm Br}}(B_s \to \mu^+\mu^-) & 
  = \left(3.0^{+0.9}_{-0.8}({\rm stat})^{+0.6}_{-0.4}({\rm sys})\right) \cdot 10^{-9}
\end{align}
and the slightly better signal significance of $4.3\,\sigma$.

The large decay width difference $\Delta \Gamma_s$ of the $B_s$ system implies
that the instantaneous branching ratio at time $t = 0$, ${\rm Br}^{[t=0]} (B_q
\to \ell^+\ell^-)$, deviates from $\overline{\rm Br}$. Neglecting for a moment
cuts on the lifetime in the experimental determination of $\overline{{\rm Br}}$,
the fully time-integrated and the instantaneous branching ratios are related in
the SM as~\cite{DeBruyn:2012wk}
\begin{equation}
  \label{eq:Br:t-integrated}
  \overline{{\rm Br}}  = \frac{{\rm Br}^{[t=0]}}{1 - y_q}\,,
  \quad
  \text{where}
  \quad
  y_q  = \frac{\Delta \Gamma_q}{2 \Gamma_q}\,.
\end{equation}
LHCb has measured $y_s = 0.088 \pm 0.014$ \cite{Raven:2012fb, LHCb:2012ad} and
established a SM-like sign for $\Delta \Gamma_s$ \cite{Aaij:2012eq}.  By 2018,
the experimental accuracy in $\overline{{\rm Br}}$ is expected to reach 
$0.5 \cdot 10^{-9}$ and with $50$ fb$^{-1}$ $0.15 \cdot 10^{-9}$ \cite{Bediaga:2012py},
the latter corresponding to the level of about 5\% error with respect to the
current central value. Results of comparable precision may be expected from CMS,
and perhaps also from ATLAS.

Motivated by the experimental prospects, this work presents a complete
calculation of the next-to-leading (NLO) electroweak (EW) matching corrections
in the SM, supplemented with the effects of the QED renormalization group
evolution (RGE).  Thereby, we remove a sizable uncertainty which has often been
neglected in the past and became one of the major theoretical uncertainties
after the considerable shrinking of hadronic uncertainties from recent progress
in lattice QCD.
  
After decoupling the heavy degrees of freedom of the SM -- the top quark, the
weak gauge bosons and the Higgs boson -- at lowest order in EW interactions, the
decay $B_q \to \ell^+\ell^-$ is governed by an effective $\Delta B = 1$
Lagrangian
\begin{align}
  \label{eq:DeltaB1:eff:Lag} 
  {\cal L}_{\rm eff} & 
   = V_{tb}^{}V_{tq}^*\, {\cal C}_{10} P_{10}
   + {\cal L}_{{\rm QCD} \times {\rm QED}}^{(5)} 
   + \mbox{h.c.}
\end{align}
with a single operator $P_{10} = [\bar{q}_L\, \gamma_\mu \, b_L ] [\bar{\ell}\,
\gamma^\mu \gamma_5\, \ell]$ and its Wilson coefficient ${\cal C}_{10}$, as well
as the QCD$\times$QED interactions of leptons and five light quark
flavors. $V_{ij}$ denotes the relevant elements of the Cabibbo-Kobayashi-Maskawa
(CKM) quark mixing matrix. Here we deviate from the usual convention to factor
out combinations of EW parameters\footnote{Since we shall not vary the EW
  renormalization scheme of the CKM factor $V_{tb}^{} V_{tq}^*$, we prefer to
  keep it as a prefactor, to have a universal ${\cal C}_{10}$ for both $q =
  d,s$.}, such as Fermi's constant, $\GF{}$, the QED fine structure constant,
$\alpha_e$, the $W$-boson mass, $M_W$, or the sine of the weak mixing angle $s_W
\equiv \sin(\theta_W)$. The most common normalizations are
\begin{align}
  \label{eq:c10:norm}
  {\cal C}_{10} & 
  = \frac{4\hspace{0.01cm} \GF{}}{\sqrt{2}}\, c_{10} \,,
&
  {\cal C}_{10} & 
  = \frac{\GF{2} \hspace{0.01cm} M_W^2}{\pi^2}\, \widetilde{c}_{10} \,,
\end{align}
with the LO Wilson coefficients
\begin{align}
  \label{eq:C10LO}
  c_{10} & = - \frac{\alpha_e}{4\pi} \frac{Y_0(x_t)}{s_W^2}\,, 
&
  \widetilde{c}_{10} & = -Y_0(x_t) \,.
\end{align}
They depend on the gauge-independent function $Y_0$ \cite{Inami:1980fz}, where
$x_t = (M_t/M_W)^2$ denotes the ratio of top-quark to $W$-boson masses. We will
frequently refer to the choice $c_{10}$ and $\widetilde{c}_{10}$ as the
``single-$\GF{}$'' and ``quadratic-$\GF{}$'' normalization, respectively. The
former choice is the standard convention of the $\Delta B = 1$ effective theory
in the literature, whereas the latter choice removes the dependence on
$\alpha_e$ and $s_W$ in favor of $\GF{}$ and $M_W$ \cite{Misiak:2011bf}.  At
lowest order in the EW interactions both normalizations may be considered
equivalent due to the tree-level relation $\GF{} = \pi \alpha_e/(\sqrt{2} M_W^2
s_W^2)$. 
In practice, however, large differences arise once numerical input for
the EW parameters is used that corresponds to different renormalization
schemes. For example, a noticeable $7$\% change of the branching ratio is
caused by choosing $s_W^2 = 0.2231$ in the on-shell scheme instead of 
$s_W^2 = 0.2314$ in the $\MS$ scheme with the numerical values taken from
Ref.~\cite{Beringer:1900zz}.  At higher orders in EW couplings, the analytic form
of ${\cal C}_{10}$ depends on the choice of normalization as well as the EW
renormalization scheme of the involved parameters. Especially the power of
$\GF{}$ affects the matching, whereas the choice of EW renormalization scheme
implies different finite counterterms for the parameters. Thereby, the overall
numerical differences among the different choices of normalizations and EW
renormalization schemes become much smaller, removing the large uncertainty
present at lowest order.

The instantaneous branching ratio takes the form
\begin{align}
  \label{eq:BR:time0}
  {\rm Br}^{[t=0]}(B_q \to \ell^+\ell^-) & =
    {\cal N} \, \big|{\cal C}_{10} \big|^2\,, 
\end{align}
with the normalization factor
\begin{align}
  {\cal N}& = 
    \frac{\tau_{B_q} M_{B_q}^3 f_{B_q}^2}{8\,\pi} |V_{tb}^{}V_{tq}^*|^2 
    \frac{m_\ell^2}{M_{B_q}^2} \,
    \sqrt{1 - 4\, m_\ell^2 / M_{B_q}^2} \,.       
\end{align}
It exhibits the helicity suppression due to the lepton mass $m_\ell$ and depends
on the lifetime $\tau_{B_q}$ and the mass $M_{B_q}$ of the $B_q$
meson. Moreover, a single hadronic parameter enters, the $B_q$ decay constant
$f_{B_q}$,
\begin{align}
  \left\langle 0 |\bar{q}\, \gamma_\mu \gamma_5\, b | \bar{B}_q(p) 
  \right \rangle & = 
    i f_{B_q} p_\mu\,.
\end{align}
It is nowadays subject to lattice calculations with errors at a few percent
level, eliminating this previously major source of uncertainty
\cite{McNeile:2011ng, Bazavov:2011aa, Na:2012kp, Dowdall:2013tga}.  The
uncertainties due to $f_{B_q}$, $\tau_{B_q}$ and $y_q$ approach a level of below
$3\%$ \cite{Buras:2013uqa} in $\overline{\rm Br}$. Concerning perturbative
uncertainties, the strong dependence of ${\cal C}_{10}$ on the choice of the
renormalization scheme for $M_t$ is removed when including the NLO QCD
contribution in the strong coupling $\alpha_s$ \cite{Buchalla:1992zm,
  Buchalla:1993bv, Misiak:1999yg, Buchalla:1998ba}. So far the full NLO EW
corrections have not been calculated and in this work we close this gap as
previously done for the analogous corrections to $s \to d \nu \bar{\nu}$
\cite{Brod:2010hi}. Being usually ignored in the budget of theoretical
uncertainties of Eq.~\eqref{eq:BR:time0}, the importance of a complete calculation
has recently been emphasized~\cite{Buras:2012ru}.  There, the NLO EW corrections
in the limit of large top-quark mass have been employed, which is known to be
insufficient at the level of accuracy aimed at Ref.~\cite{Brod:2010hi} and the
residual EW uncertainties were estimated to be at least $5\%$ on the branching
ratio.

In \refsec{sec:matching} we describe the calculation of the NLO~EW correction to
${\cal C}_{10}$ adopting the two choices of normalization and using different
renormalization schemes for the involved EW parameters. In \refsec{sec:RGE}, we
summarize the solution of the RGE and obtain ${\cal C}_{10}$ at the low-energy
scale of the order of the bottom-quark mass at the NLO in EW interactions. 
Finally, in \refsec{sec:num:analysis} we discuss the reduction of
the EW renormalization-scheme dependences in ${\cal C}_{10}$ after the inclusion
of NLO EW corrections. In the accompanying appendices \ref{app:technicaldetails}
and \ref{app:RGE} we collect additional technical information on the matching
calculation and the RGE, respectively. Some supplementary details of
\refsec{sec:num:analysis} have been relegated to \refapp{app:OS-1:numerics}.

%
%
\section{Matching Calculation of NLO Electroweak Corrections
  \label{sec:matching}}

We obtain the EW NLO corrections to the Wilson coefficient ${\cal C}_{10}$ by
matching the effective theory of EW interactions to the Standard Model.  For
this purpose we evaluate one-light-particle irreducible Greens functions with
the relevant external light degrees of freedom up to the required order in the
EW couplings in both theories. The Wilson coefficients are determined by
requiring equality of the renormalized Greens functions order by order
\begin{equation}
  \label{eq:matchingamplitudes}
  {\cal A}_{\rm full}(\mu_0) \overset{\underset{!}{}}{=}
  {\cal A}_{\rm eff}(\mu_0) \,
\end{equation}
at the matching scale $\mu_0$. It is chosen of the order of the masses of the
heavy degrees of freedom to minimize otherwise large logarithms that enter the
Wilson coefficients. The Wilson coefficients have the general expansion
\begin{equation}
  \label{eq:WC:matching:scale}
\begin{aligned}
  {\cal C}_i(\mu_0) & = 
    {\cal C}_i^{(00)} + \tilde{\alpha}_s\, {\cal C}_i^{(10)} 
    + \tilde{\alpha}_s^2\, {\cal C}_i^{(20)}
\\[0.2cm]
  & \quad
  + \tilde{\alpha}_e \left({\cal C}_i^{(11)} + \tilde{\alpha}_s\, {\cal C}_i^{(21)}
     + \tilde{\alpha}_e\, {\cal C}_i^{(22)}\right) 
  + \ldots \,,
\end{aligned}
\end{equation}
in the strong and electromagnetic $\tilde{\alpha}_{s,e} \equiv
\alpha_{s,e}/(4\pi)$ running couplings of the effective theory at the scale
$\mu_0$, where we follow the convention of Ref.~\cite{Huber:2005ig}. This expansion
starts with tree-level contributions denoted by the superscript $(00)$, has
higher-order QCD corrections $(m0)$ with $m>0$, pure QED corrections $(mm)$ with
$m>0$ and mixed QCD-QED corrections $(mn)$ with $m>n>0$, all of which depend
explicitly on $\mu_0$ except for $(00)$. For ${\cal C}_{10}$ the non-zero
matching corrections start at order $\tilde{\alpha}_e$, i.e., for $n\geq 1$. The
${\cal C}_{10}^{(11)}$ \cite{Inami:1980fz} and ${\cal C}_{10}^{(21)}$ 
\cite{Buchalla:1992zm, Buchalla:1993bv, Misiak:1999yg, Buchalla:1998ba} contributions
are known and here we calculate ${\cal C}_{10}^{(22)}$. Above, 
Eq.~\eqref{eq:WC:matching:scale} has to
be understood as the definition of the components ${\cal C}_i^{(mn)}$ that
complies with the single-$\GF{}$ normalization in the literature
\cite{Huber:2005ig}. Comparison with Eqs.~\eqref{eq:c10:norm} and
\eqref{eq:C10LO} yields
\begin{align}
  {\cal C}_{10}^{(11)} &
  = \frac{4 \GF{}}{\sqrt{2}} c_{10}^{(11)} 
  = - \frac{4 \GF{}}{\sqrt{2}} \frac{Y_0(x_t)}{s_W^2}
\intertext{and}
  {\cal C}_{10}^{(11)} &
  = \frac{\GF{2} M_W^2}{\pi^2 \tilde{\alpha}_e} \widetilde{c}_{10}^{\,(11)} 
  = - \frac{\GF{2} M_W^2}{\pi^2 \tilde{\alpha}_e} Y_0(x_t)
\end{align}
showing that this convention introduces an artificial factor $1/\alpha_e$ into
the components in the case of the quadratic-$\GF{}$ normalization. However, we
will organize the renormalization group evolution (see \refsec{sec:RGE}) such
that these factors are of no consequence, as should be.

Although the operator $P_{10}$ does not mix with other $\Delta B = 1$ operators
under QCD, at higher order in QED interactions such a mixing does take place
\cite{Bobeth:2003at, Huber:2005ig}. As a consequence the effective Lagrangian
\eqref{eq:DeltaB1:eff:Lag} has to be extended
\begin{align}
  \label{eq:DeltaB1:eff:Lag:2}
  {\cal C}_{10} P_{10} & \quad \longrightarrow \quad \sum_i {\cal C}_i P_i\,,
\end{align}
where the term $\sim V_{ub}^{}V_{uq}^* \,[{\cal C}_1^{} (P_1^c - P_1^u) 
+ {\cal C}_2^{} (P_2^c - P_2^u)]$ does not contribute to the order considered
here. 
The operators relevant to $B_q \to \ell^+\ell^-$ at the considered order in
strong and EW interactions comprise the current-current operators ($i = 1,2$),
QCD-penguin operators ($i = 3,4,5,6$) and the semi-leptonic operator
($i = 9, 10$). We follow the operator definition of Ref.~\cite{Huber:2005ig}
that does not include the factor $\alpha_e/(4 \pi)$ in $P_{9,10}$. This
factor is included in the matching conditions of the Wilson coefficients at
the matching scale $\mu_0$ in Eq.~\eqref{eq:WC:matching:scale}. In the matching
calculation only $P_2$ and $P_9$ as defined in \refapp{app:operatorbasis}
are needed, whereas the remaining operators enter in the renormalization
group evolution discussed in \refsec{sec:RGE}.

We describe the calculation of ${\cal A}_{\rm full}$ and ${\cal A}_{\rm eff}$ in
Sections~\ref{sec:fullcalc} and \ref{sec:effcalc}, respectively. In the SM
calculation of ${\cal A}_{\rm full}$, we apply different EW renormalization
schemes for the involved parameters to demonstrate in \refsec{sec:num:analysis}
that the renormalization scheme dependence 
is reduced to sub-percent effects
when including ${\cal C}_{10}^{(22)}$. The schemes differ by finite parts of the counterterms
that renormalize the bare parameters of the Lagrangian or equivalently the
parameters appearing in the LO Wilson coefficient.  Nevertheless, we use the
same physical input in all schemes for the numerical evaluation that we have
chosen to be
\begin{equation}
  \label{eq:EWinput}
\begin{aligned}
   & \GF{}, \quad \alpha_e(M^{\rm pole}_Z), \quad \alpha_s(M^{\rm pole}_Z),  
\\[0.2cm]
   & V_{ij}, 
   \quad M^{\rm pole}_Z, \quad M^{\rm pole}_t, \quad M^{\rm pole}_H\,. 
\end{aligned}
\end{equation}
$\GF{}$ is the Fermi constant as extracted from muon life-time experiments.  It
is itself a Wilson coefficient of the effective theory and plays thus a special
role in the calculation of EW corrections; we postpone further discussion to
Section~\ref{sec:effcalc}. The couplings $\alpha_e$ and $\alpha_s$ are the $\MS$
couplings at the scale of the $Z$ pole mass in the SM with decoupled top quark\footnote{
I.e. $W$ and $Z$ bosons are still dynamical degrees of freedom.}.
$V_{ij}$ are elements of the CKM matrix.  $M^{\rm pole}_Z$, $M^{\rm
  pole}_t$ and $M^{\rm pole}_H$ are the pole masses of $Z$ boson, top quark and
Higgs boson, respectively.  The numerical values are summarized in
\reftab{tab:numinput}.

\begin{table}
\begin{center}
\renewcommand{\arraystretch}{1.2}
\begin{tabular}{lll}
  {\bf Parameter}	
& {\bf Value}
& {\bf Ref.}
\\
\hline
  $\GF{}$
& $1.166\,379\, \cdot 10^{-5}\, \rm{GeV}^{-2}$
& \cite{Beringer:1900zz}	
\\ 
  $\alpha_s^{}(M_Z^{\rm pole})$\quad$(N_f=5)$	
& $0.1184 \pm 0.0007$		
& \cite{Beringer:1900zz}	
\\
  $\alpha_e^{}(M_Z^{\rm pole})$\quad$(N_f=5)$	
& $(127.944 \pm 0.014)^{-1}$
& \cite{Beringer:1900zz}				
\\
  $M_Z^{\rm pole}$				
& $(91.1876 \pm 0.0021)$ GeV
& \cite{Beringer:1900zz}
\\
  $M_t^{\rm pole}$
& $(173.1 \pm 0.9)$ GeV
& \cite{Beringer:1900zz, Lancaster:2011wr, Aaltonen:2012ra}
\\
  $M_H^{\rm pole}$
& $(125.9 \pm 0.4)$ GeV
& \cite{Beringer:1900zz, Aad:2012tfa, Chatrchyan:2012ufa}
\\
  $\Delta \alpha_{e,{\rm hadr}}^{\rm (5)}(M_Z^{\rm pole})$
& $0.02772\pm 0.00010$
& \cite{Beringer:1900zz}
\\
\hline
\end{tabular}
\caption{
  \label{tab:numinput}
  The physical input. $\alpha_{s,e}$ are the running $\MS$ couplings of the
  five-flavor theory at $\mu=M_Z$. Masses are the experimentally measured pole
  masses.}
\renewcommand{\arraystretch}{1.0}
\end{center}
\end{table}

%
\subsection{Standard Model Calculation \label{sec:fullcalc}}

We keep only the leading contributions of the expansion in the momenta of
external states, in which case the full amplitude for $b\to q\ell^+\ell^-$ takes
the form
\begin{align}
  \label{eq:afull}
  {\cal A}_{\rm full} & 
  = \sum_i A_{{\rm full},\, i}(\mu) \langle P_i(\mu)\rangle^{(0)}\,.
\end{align}
$\langle P_i(\mu)\rangle^{(0)}$ denote the tree-level matrix elements of
operators mediating $b\to q\ell^+\ell^-$, i.e., $i =9, 10$ as well as evanescent
operators defined in \refapp{app:operatorbasis}.  The $A_{{\rm full}, i}$'s are
coefficient functions with the electroweak expansion
\begin{align}
  \label{eq:afull:series}
  A_{{\rm full},\,i} & 
  = A_{{\rm full},\,i}^{(0)} 
  + \tilde\alpha_e A_{{\rm full},\,i}^{(1)} 
  + \tilde\alpha_e^2 A_{{\rm full},\,i}^{(2)}
  +\dots\,,
\end{align}
with $\alpha_e$ of the SM, i.e. six active quark flavors as well as heavy weak
gauge bosons and the Higgs boson. In the case of the single-$\GF{}$
normalization, $A_{{\rm full},\,i}^{(0)} = 0$ for $b\to q\ell^+\ell^-$ mediating
operators, whereas $A_{{\rm full},\,i}^{(0)} \neq 0$ for the quadratic-$\GF{}$
normalization due to the substitution $\alpha_e/s_W^2 \to \GF{}$.

\begin{figure*}[]
\begin{center}
\includegraphics[]{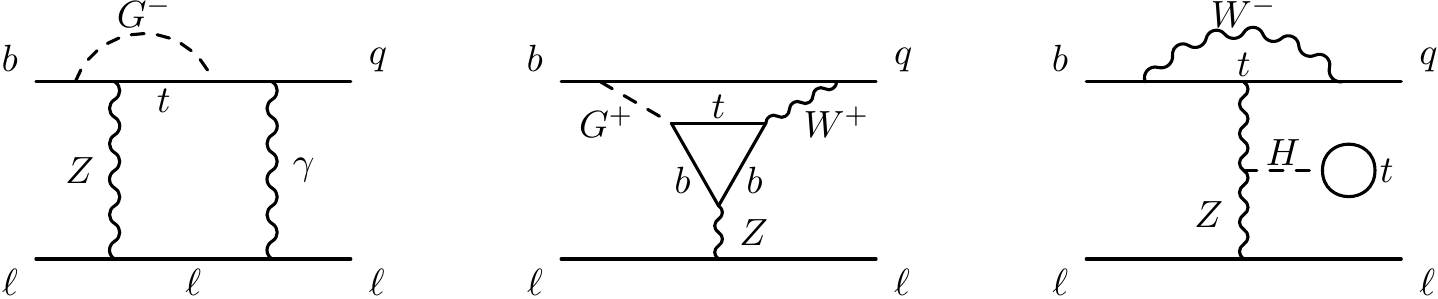}
\end{center}
\caption{Two-loop diagrams in the SM contributing to the $b\to q\ell^+\ell^-$ at
NLO in EW interactions.}
\label{fig:SMdiagrams}
\end{figure*}

Our focus here is the calculation of the two-loop contribution to $A_{{\rm full},
 10}$ and some parts of $A_{{\rm full},\,i}$ at one-loop that
involve evanescent operators $E_9$ and $E_{10}$ (see
\refapp{app:operatorbasis}).  For this purpose, we calculate all two-loop EW
Feynman diagrams and the corresponding one-loop diagrams with inserted
counterterms, \reffig{fig:SMdiagrams} depicts some examples. 
We proceed as in Ref.~\cite{Brod:2010hi} and perform all calculations in the Feynman gauge $\xi = 1$
using two independent setups. 
Similarly to Ref.~\cite{Brod:2010hi} also here we find contributions from electroweak 
gauge bosons that are $1/s_W^2$ enhanced. 
In \refapp{app:fullcalc} we discuss the more
technical aspects of the calculation, e.g.~$\gamma$-algebra in $d$-dimensions
and loop-integrals. 
Here, we concentrate on the electroweak renormalization
conditions.

Having fixed the physical input, we define three renormalization schemes and
discuss the relation of their renormalized parameters to the physical input in
Eq.~\eqref{eq:EWinput}. In all three schemes we use $\MS$ renormalization for
$\alpha_e$ and the top-quark mass under QCD, whereas additional finite terms are
included into the field renormalization constants as explained in more detail in
\refapp{app:fullcalc}.  Therefore, our schemes differ only by finite EW
renormalizations of $s_W$, $M_t$ and $M_W$ appearing at LO in $c_{10}$. For
$\widetilde c_{10}$, $s_W$ is absorbed in the additional factor $\GF{}$ and
needs no further specification. \\[-0.2cm]

{\bf 1.) On-shell scheme}

In the on-shell scheme, at the order we consider, the on-shell masses of $Z$
boson and top quark coincide with their pole masses. The mass of the $W$ boson is
a dependent quantity for our choice of physical input.  We calculate it
including radiative corrections following Ref.~\cite{Awramik:2003rn}. This relation
introduces a mild Higgs-mass dependence of ${\cal C}_{10}$ at LO. The weak
mixing angle in the on-shell scheme is defined by
\begin{align}
  s_W^2 & 
  \equiv (s_W^\OS)^2 
  = 1 -\left(M_W^{\OS}/M_Z^\OS\right)^2\,.
  \label{eq:swonshell}
\end{align}
Therefore, the only finite counterterms necessary are $\delta M_Z^2$, $\delta
M_W^2$ and $\delta M_t$ at one-loop, they are given in Refs.~\cite{Jegerlehner:2001fb,
  Jegerlehner:2002em}. We also treat tadpoles as in Refs.~\cite{Jegerlehner:2001fb,
  Jegerlehner:2002em}: we include tadpole diagrams (see
Fig.~\ref{fig:SMdiagrams}), and a renormalization $\delta t$ to cancel the
divergence and the finite part of the one-loop tadpole diagram. This way we
ensure that all renormalization constants apart from wave function
renormalizations are gauge invariant
\cite{Fleischer:1980ub}. \\[-0.2cm]

{\bf 2.) $\boldsymbol\MS$ scheme}

In the $\MS$ scheme the fundamental parameters are those of the ``unbroken'' SM
Lagrangian
\begin{align}
  \label{eq:msbarinput}
  g_1,&& g_2,&& g_3,&& v,&& \lambda && \mbox{and} &&y_t.
\end{align}
Here $g_3$, $g_2$ and $g_1$ are the couplings of the SM gauge group $SU(3)_c
\times SU(2)_L\times U(1)_Y$, $v$ is the vacuum expectation value of the Higgs
field and $\lambda$ its quartic self-coupling, whereas $y_t$ is the top-Yukawa
coupling. The parameters are renormalized by counterterms subtracting only
divergences and $\log(4\pi) -\gamma_E$ terms, i.e., they are running $\MS$
parameters. We do not treat tadpoles differently in this respect, only their
divergences are subtracted by the counterterm for $v$. By expressing the
parameters of the LO Wilson coefficients in terms of the ``unbroken''-phase
parameters
\begin{equation}
  \label{eq:msbarcouplings}
\begin{aligned}
  s_W^2 & = g_1^2/(g_1^2+g_2^2)\,,	&	
  4 \pi \alpha_e & = g_1^2 g_2^2/(g_1^2+g_2^2)\,,&
\\
  M_W & = v g_2/2 \,, &
  x_t & = 2 y_t^2/g_2^2\,,&
\end{aligned}
\end{equation}
we iteratively fix the values of the ``unbroken'' parameters at the matching
scale $\mu_0$. To this end, we require that the physical input in Eq.~\eqref{eq:EWinput}
is reproduced to one-loop accuracy. \\[-0.2cm]

{\bf 3.) Hybrid scheme}

For Eq.~\eqref{eq:C10LO}, where $s_W$ appears at LO, we may adopt yet another
scheme. We renormalize the couplings $\alpha_e$ and $s_W$ in the $\MS$ scheme
and the masses in $x_t$ on-shell. Effectively this corresponds to including the
on-shell counterterms for masses and using Eq.~\eqref{eq:msbarcouplings} instead
of Eq.~\eqref{eq:swonshell} for $s_W$. Correspondingly, we use $s_W$, $\alpha_e$,
$M_t$, $M_W$ and $M_H$ as fundamental parameters for the hybrid scheme. This
scheme is a better-behaved alternative to the on-shell scheme, in which the
counterterm for $s_W$ receives large top-quark mass dependent corrections.
(see App.~\ref{app:OS-1:numerics}).

Having fixed all renormalization conditions we evaluate $A_{{\rm full},10}^{(2)}$.
In practice we calculate the $\MS$ amplitude and include the
appropriate counterterms in $A_{{\rm full}, 10}^{(1)}$ to shift from the
$\MS$ to the on-shell or hybrid scheme. The full expression for $A_{{\rm
    full}, 10}^{(2)}$ is too lengthy to be included here\footnote{We
  attach the complete analytic two-loop EW contribution in the on-shell scheme
  for the quadratic-$\GF{}$ normalization, $\widetilde{c}_{10}^{\,(22)}$, to the
   electronic preprint.}.

%
\subsection{Effective Theory Calculation \label{sec:effcalc}}

The effective theory is described by the effective Lagrangian in
Eqs.~\eqref{eq:DeltaB1:eff:Lag} and \eqref{eq:DeltaB1:eff:Lag:2} with
canonically normalized kinetic terms for all fields. To simplify the notation we
drop any indices indicating an expansion in $\tilde \alpha_s$ throughout this
Section.  The fields and couplings are $\MS$-renormalized via the redefinitions
of bare quantities
\begin{equation}
\begin{aligned}
  d & \to \sqrt{Z_{d}}\, d\,, &
  \ell & \to \sqrt{Z_{\ell}}\,\ell\,, &
  C_{j} & \to \sum_{i}C_i\, \hat Z_{i,j} \,, &
\end{aligned}
\end{equation}
where $d$ denotes down-type quark fields and $\ell$ denotes charged-lepton fields.
The renormalization constant of the Wilson coefficients is the matrix $\hat Z_{i,j}$
arising from operator mixing. It has an expansion in $\tilde\alpha_e$
\begin{align}
  \label{eq:RC-expansion}
  \hat Z_{i,j} & 
   = \delta_{i,j} + \tilde \alpha_e\, \hat Z_{i,j}^{(1)} 
   + \tilde \alpha_e^2\, \hat Z_{i,j}^{(2)} + \dots
\end{align} 
analogously to the expansion of the renormalization constants of 
the fields and couplings given in Eq.~\eqref{eq:eff-th:ren-constants}.

All loop diagrams in the effective theory vanish, since we set all
light masses to zero, expand in external momenta and employ
dimensional regularization. Accordingly, the effective theory amplitude
is entirely determined through the product of tree-level matrix
elements $\langle P_j \rangle^{(0)}$,
the Wilson coefficients ${\cal C}_i$ and appropriate
renormalization constants. 
The renormalized amplitude reads
\begin{equation}
  \label{eq:eff-th:amp}
\begin{split}
  {\cal A}_{\rm eff}(\mu) 
  & = \sum_{i} A_{{\rm eff},\, i}(\mu) \langle P_i(\mu) \rangle^{(0)}\\
  & = V_{tb}^{} V_{tq}^* \sum_{i,j} 
      {\cal C}^{}_i(\mu) \, \hat Z_{i,j} Z_j \, \langle P_j(\mu) \rangle^{(0)}\,.
\end{split}
\end{equation}
As mentioned above, both the Wilson coefficients ${\cal C}_i$ and the
renormalization constants are expanded in $\tilde{\alpha}_{e}$ as given in
Eqs. \eqref{eq:WC:matching:scale} and \eqref{eq:RC-expansion}, respectively. The
$Z_j$'s summarize products of field- and charge-renormalization constants of the
operator in question, i.e. for $P_{10}$
\begin{align}
  Z_{10} & = Z_{d}\, Z_{\ell}\,,
\end{align}
which is the one required up to two-loop level in $\tilde\alpha_e$.

Only a few physical operators contribute to the part of the amplitude
in Eq.~\eqref{eq:eff-th:amp} proportional to $\langle P_{10} \rangle^{(0)}$
since only a few mix either at one-loop or two-loop level into $P_{10}$ and have,
at the same time, a non-zero Wilson coefficient at one-loop or tree-level,
respectively. These are: the operator $P_2$ having a non-zero Wilson coefficient
${\cal C}_2^{(00)}$ as well as an entry in $\hat{Z}_{2,10}^{(2)}$ and $P_9$
that mixes
at one-loop into $P_{10}$ and have a non-vanishing~${\cal
  C}_i^{(11)}$.
Apart from the physical operators also 
one evanescent operator, i.e.\ $E_9$ contributes.
We give the definition of the operators in
\refapp{app:operatorbasis} and present some details on the calculation of the
renormalization constants in the five-flavor theory in \refapp{app:effcalc}. All
contributing mixing renormalization constants of physical operators can be
extracted from the anomalous dimension in the literature
\cite{Bobeth:2003at}. We collect all constants and discuss the mixing of
evanescent operators in App.~\ref{app:effcalc}. Finally, at the two-loop level
\begin{equation}
  \label{eq:A2eff}
\begin{aligned}
  A_{{\rm eff},10}^{(2)} & = 
   V_{tb}^{} V_{tq}^* \, (\tilde\alpha_e)^{n} \biggl[
   {\cal C}_{10}^{(22)}
   + {\cal C}_{10}^{(11)} Z^{(1)}_{10}
\\[0.5em]&
   + {\cal C}_{2}^{(00)} \hat{Z}^{(2)}_{2, 10}
   + \sum_{i = 9, E_9}
     {\cal C}_{i}^{(11)} \hat Z^{(1)}_{i, 10}
   \biggl] \,
\end{aligned}
\end{equation}
with the power $n=2$ and $n=1$ for the single- and quadratic-$\GF{}$ normalization,
respectively. In this equation $\alpha_e$ is the electromagnetic
coupling constant in the $\Delta B = 1$ effective theory. It differs from
the one in \reftab{tab:numinput} by threshold corrections due to $W$ and $Z$ gauge
bosons and from the one in the SM in Eq.~\eqref{eq:afull:series} by the additional
top-quark threshold corrections as explained above Eq.~\eqref{eq:alpha_e:threshold}. 
Note that the renormalization constant $\hat{Z}^{(2)}_{2, 10}$, see Eq.~\eqref{eq:Z2:210},
implies the existence of a quadratic logarithm that will be resummed
with the help of the RGE 
in \refsec{sec:RGE}.
 
The one-loop Wilson coefficients in Eq.~\eqref{eq:A2eff}, multiplied with
renormalization constants, contribute finite terms to the matching through their
${\cal O}(\epsilon)$ terms. We reproduce the finite and ${\cal O}(\epsilon)$
parts of ${\cal C}_{9, 10}^{(11)}$ in \cite{Bobeth:1999mk}. For ${\cal
  C}_{E_{9}}^{(11)}$ only the finite term is needed, we give it in
App.~\ref{app:effcalc}. For this purpose we have matched also the one-loop
amplitudes proportional to the $\langle P_{9,\, 10},\, E_{9}
\rangle^{(0)}$ keeping ${\cal O}(\epsilon)$ terms when required.

The Fermi constant, $\GF{}$, is very precisely measured in muon decay and
provides a valuable input for the determination of the EW parameters. Following
\cite{Brod:2010hi}, we define $\GF{}$ to be proportional to the Wilson
coefficient $G_\mu$ of the operator $Q_\mu = (\bar\nu_{\mu_L} \gamma_\rho
\mu_L)(\bar e_L \gamma^\rho \nu_{e_L})$ that induces muon decay in the effective
Fermi theory
\begin{equation}
  \label{eq:Gmu}
  \GF{} \equiv \frac{1}{2\sqrt{2}} G_\mu 
  = \frac{1}{2\sqrt{2}}  \left(G_\mu^{(0)}  
  + \tilde\alpha_e G_\mu^{(1)} + \dots\right)\,,
\end{equation}
with the tree-level matching relation 
\begin{align}
  G_\mu^{(0)} & 
  = \frac{2\pi\alpha_e}{s_W^2 M_W^2} 
  = \frac{2}{v^2} 
\end{align}
and the NLO EW correction $G_\mu^{(1)}$. Since we work at NLO in EW interactions,
$G_\mu^{(1)}$ enters the effective theory amplitude in Eq.~\eqref{eq:eff-th:amp}.
Moreover, the power of $\GF{}$ in the normalization of the effective Lagrangian
affects the matching contribution of $G_\mu^{(1)}/G_\mu^{(0)} \times {\cal
  C}_i^{(11)}$ to ${\cal C}_i^{(22)}$, in contrast to the leading EW components
${\cal C}_i^{(11)}$ that remain unchanged when using different powers. This can
be best understood by the explicit $\tilde\alpha_e$ expansion for the
single-$\GF{}$ normalization
\begin{align}
  {\cal C}_{10} & 
  \sim \GF{} \, c_{10}
  \sim \big[G_\mu^{(0)} + \tilde\alpha_e G_\mu^{(1)} \big]
    \big[c_{10}^{(11)} + \tilde{\alpha}_e\, c_{10}^{(22)}\big]
\\[0.2cm] 
  \nonumber
& = G_\mu^{(0)} \left[ c_{10}^{(11)} 
    + \tilde\alpha_e \left( c_{10}^{(22)} 
       +  \frac{G_\mu^{(1)}}{G_\mu^{(0)}}\, c_{10}^{(11)}  \right)
  \right]
  + {\cal O}(\tilde{\alpha}_e^2)
\intertext{and the quadratic-$\GF{}$ normalization}
  {\cal C}_{10} & \sim
    (G_\mu^{(0)})^2 \left[ \widetilde{c}_{10}^{\,(11)} 
      + \tilde\alpha_e \left( \widetilde{c}_{10}^{\,(22)} 
         + 2 \, \frac{G_\mu^{(1)}}{G_\mu^{(0)}}\, \widetilde{c}_{10}^{\,(11)} \right)
  \right]\,,
\end{align}
which receives an additional factor of 2. Depending on the choice of
normalization, the according contribution proportional to $G_\mu^{(1)}/
G_\mu^{(0)} \times {\cal C}_i^{(11)}$ enters Eq.~\eqref{eq:A2eff}.

The merit of defining $\GF{}$ to be itself a Wilson coefficient at the matching
scale is that the large uncertainties from the scale dependence of the vacuum
expectation value in $G_\mu^{(0)}$ do not appear at all at LO in the Wilson
coefficient.

This way, we obtain ${\cal C}_{10}^{(22)}$, which has been known only in the
large top-quark-mass limit \cite{Buchalla:1997kz, Fleischer:1994cb}, by matching
the parts of $A_{\rm eff} \sim \langle P_{10} \rangle^{(0)}$ and $A_{\rm full}
\sim \langle P_{10} \rangle^{(0)}$ at NLO order in $\tilde{\alpha}_e$ and verify
the explicit cancellations of all left-over divergences.

%
%
\section{Renormalization Group Evolution \label{sec:RGE}}

This section summarizes the results of the evolution of the Wilson coefficients
under the renormalization group equations from the matching scale $\mu_0$ down
to the low scale $\mu_b$. The matching scale $\mu_0$ is of the order of the
masses of the decoupled heavy degrees of freedom $\sim 100$~GeV and $\mu_b \sim
5$~GeV of the order of the bottom-quark mass at which matrix elements are 
evaluated. The according anomalous dimension matrices of the $\Delta B = 1$
effective theory, including NLO EW corrections, are given in
Ref.~\cite{Bobeth:2003at} and the RGE is solved in Ref.~\cite{Huber:2005ig} for the
single-$\GF{}$ normalized Lagrangian in Eqs.~\eqref{eq:DeltaB1:eff:Lag} and
\eqref{eq:C10LO} including the running of $\alpha_e$. These corrections have
already been considered in Ref.~\cite{Misiak:2011bf} in the analysis of 
$B_q \to \ell^+\ell^-$.

The evolution operator $U(\mu_b,\mu_0)$ relates the Wilson coefficients at the
matching scale, see Eq.~\eqref{eq:WC:matching:scale}, to the ones at $\mu_b$:
\begin{align}
  \label{eq:RGE:general:sol}
  {\cal C}_{i}(\mu_b) & = \sum_j U(\mu_b, \mu_0)_{ij} \,{\cal C}_{j}(\mu_0) \,.
\end{align}
At the low-energy scale the Wilson coefficients may again be expanded in
$\alpha_s(\mu_b)$ and the small ratio $\kappa \equiv \alpha_e(\mu_b) /
\alpha_s(\mu_b)$:
\begin{align}
  \label{eq:WC:low:scale}
  {\cal C}_i(\mu_b) & = 
    \sum_{m,n=0} \left[\tilde{\alpha}_s(\mu_b)\right]^m 
    \left[\kappa(\mu_b)\right]^n {\cal C}_{i,(mn)}\,.
\end{align}
We obtain the explicit expressions for the components ${\cal C}_{i,(mn)}(\mu_b)$
from the solution given in Ref.~\cite{Huber:2005ig} with further details and the
solution for $i=10$ presented in \refapp{app:RGE}.

In the single-$\GF{}$ normalization the Wilson coefficient $c_{10}(\mu_b)$ starts
at order $\alpha_e$ with the following non-zero contributions
\begin{equation}
\begin{aligned}
  c_{10}(\mu_b) & = 
    \tilde{\alpha}_e \left(c_{10,(11)} + \tilde{\alpha}_s c_{10,(21)} \right)
\\[0.2cm] 
   & 
   + \tilde{\alpha}_e^2 \left(
     \frac{c_{10,(02)}}{\tilde{\alpha}_s^2} 
   + \frac{c_{10,(12)}}{\tilde{\alpha}_s} 
   + c_{10,(22)}
         \right) \,.
\end{aligned}
\end{equation}
The components $c_{i,(mn)}$ are functions of the ratio $\eta \equiv \alpha_s
(\mu_0)/\alpha_s(\mu_b)$ and the high-scale components $c_j^{(mn)}$ of 
Eq.~\eqref{eq:WC:matching:scale}. For illustration, we give here numerical results
for the exemplary values $\mu_0 = 160$ GeV and $\mu_b = 5$ GeV, yielding
$\eta = 0.509$, 
\begin{equation}
  \label{eq:c10-scaled:exp}
\begin{aligned}
  c_{10,(11)} & = 
    c_{10}^{(11)} \,,
\\[0.2cm]
  c_{10,(21)} & = 
    \eta\, c_{10}^{(21)} \,,
\\[0.2cm]
  c_{10,(02)} & = 
    0.0058\, c_{2}^{(00)} \,,
\\[0.2cm]
  c_{10,(12)} & = 
    0.068\, c_{2}^{(00)} 
  + 0.005\, c_{1}^{(10)} 
  - 0.005\, c_{4}^{(10)}
\\ &
  + 0.252\, c_{9}^{(11)} 
  + 1.118\, c_{10}^{(11)} \,,
\\[0.2cm]
  c_{10,(22)} & = 
    0.133\, c_{1}^{(10)}
  + 0.066\, c_{4}^{(10)}
\\ &
  + 0.002\, c_{1}^{(20)}
  + 0.001\, c_{2}^{(20)}
  + 0.004\, c_{3}^{(20)} 
\\ &
  - 0.002\, c_{4}^{(20)}
  + 0.033\, c_{5}^{(20)}
  - 0.039\, c_{6}^{(20)}
\\ &
  - 1.593\, c_{9}^{(11)}
  - 2.226\, c_{10}^{(11)} 
\\ &
  + 0.128\, c_{9}^{(21)}
  + 0.569\, c_{10}^{(21)}
  + c_{10}^{(22)} \,.
\end{aligned}
\end{equation}
We give the explicit solution for arbitrary values of $\eta$ in
\refapp{app:RGE:sol}.  Furthermore, the $c_{10,(mn)}$ depend on the initial
matching conditions of the Wilson coefficients, the $c_i^{(mn)}$ in
Eq.\eqref{eq:WC:matching:scale}, at various orders: tree-level for $i = 2$,
one-loop in $\alpha_s$ for $i = 1, 4$ and in $\alpha_e$ for $9, 10$ and two-loop
in $\alpha_s^2$ for $i = 1, \ldots, 6$ and in $\alpha_e\alpha_s$ for $i = 9, 10$
\cite{Bobeth:1999mk} as well as the two-loop NLO EW correction for $i = 10$
presented in \refsec{sec:matching}.

We derive the equivalent expressions for the case of the quadratic-$\GF{}$
normalization from the single-$\GF{}$ normalization in Eq.~\eqref{eq:WC:low:scale}
\begin{align}
  \label{eq:WWC:low:scale}
  \widetilde c_i(\mu_b) & = 
    \sum_{m,n=0} \left[\tilde{\alpha}_s(\mu_b)\right]^{m-1} 
    \left[\kappa(\mu_b)\right]^{n-1} \widetilde c_{i,(mn)}\,.
\end{align}
For $i=10$ the lowest-order non-zero terms 
\begin{equation}
\begin{aligned}
  \widetilde{c}_{10}(\mu_b) & = 
    \widetilde{c}_{10,(11)} + \tilde{\alpha}_s \widetilde{c}_{10,(21)}
\\[0.2cm]
   & 
   + \tilde{\alpha}_e \left(
     \frac{\widetilde{c}_{10,(02)}}{\tilde{\alpha}_s^2} 
   + \frac{\widetilde{c}_{10,(12)}}{\tilde{\alpha}_s} 
   + \widetilde{c}_{10,(22)}
   \right)\,,
\end{aligned}
\end{equation}
already start at order $\alpha_e^0$. The components of the initial Wilson
coefficients in Eq.~\eqref{eq:WC:matching:scale} are related as
\begin{align}
  \widetilde c_{i}^{\,(mn)} & = s_W^2\, c_{i}^{(mn)}&
  \mbox{for}\quad 
  n < 2\,,
\end{align}
where a factor $\tilde\alpha_e(\mu_0)$ has been pulled out and substituted by
$\tilde\alpha_e(\mu_b)$. For cases $n\geq2$, which is here only of concern for
${\cal C}_{10}$, an additional shift has to be taken into account explicitly in
the matching analogously to the discussion below Eq.~\eqref{eq:Gmu}.  Eventually,
the downscaled components $\widetilde{c}_{i,(mn)}$ in Eq.~\eqref{eq:WWC:low:scale} are
given by Eq.~\eqref{eq:c10-scaled:exp} with the replacement $c_i^{(mn)} \to
\widetilde{c}_i^{\,(mn)}$ and by omitting the contributions of
$\widetilde{c}_{10}^{\,(11)}$ in $\widetilde{c}_{10,(12)}$ as well as
$\widetilde{c}_{10}^{\,(11)}$ and $\widetilde{c}_{10}^{\,(21)}$ in
$\widetilde{c}_{10,(22)}$, as explained in more detail in \refapp{app:RGE}.

%
%
\section{Numerical Impact of NLO EW Corrections \label{sec:num:analysis}}

In \refsec{sec:matching} we presented the details of the calculation of the
complete NLO EW matching corrections to the Wilson coefficient ${\cal C}_{10}$
in the SM and in \refsec{sec:RGE} the effects of the renormalization group
evolution within the $\Delta B = 1$ effective theory
from the matching scale $\mu_0$ to the low energy
scale $\mu_b$. In this section, we discuss the numerical impact of these
corrections on ${\cal C}_{10}$ at both scales and assess the reduction of
theoretical uncertainties associated with the different choices of the
renormalization scheme. Finally, we shall briefly comment on the branching ratio
${\rm Br} \propto |{\cal C}_{10}|^2$.

Throughout, we use the four-loop $\beta$ function for $\alpha_s$ including the
three-loop mixed QCD$\times$QED term given in Ref.~\cite{Huber:2005ig}.  When
crossing the $N_f = 5$ to $N_f = 6$ threshold at the matching scale $\mu_0$, we
include the three-loop QCD threshold corrections using the pole-mass value for
the top-quark mass $M_t^{\rm pole}$ (see \reftab{tab:numinput}).  The running of
$\alpha_e$ is implemented including the two-loop QED and three-loop mixed
QED$\times$QCD terms presented in \cite{Huber:2005ig}, where the threshold
corrections have been omitted when crossing the $N_f = 5$ to $N_f = 6$
threshold entering the evolution of $\alpha_s$. We list the initial
conditions for the coupling constants in \reftab{tab:numinput} and remark that
the value of $\alpha_e$ given in Ref.~\cite{Beringer:1900zz} refers to the coupling
within the SM with the top quark decoupled. From this value we determine $\alpha_e$ 
at $\mu = M_Z$ in the SM with $N_f=6$ with the help of the decoupling
relation of Eq.~\eqref{eq:alpha_e:threshold} thereby omitting the constant and logarithmic
term from the gauge boson contribution and determine the dependent EW parameters
as described in \refsec{sec:fullcalc}. The value of $\alpha_e$ in the effective
theory is found as described below the decoupling relation of Eq.~\eqref{eq:alpha_e:threshold}.

\begin{figure*}[tbp]
\begin{center}
\includegraphics[]{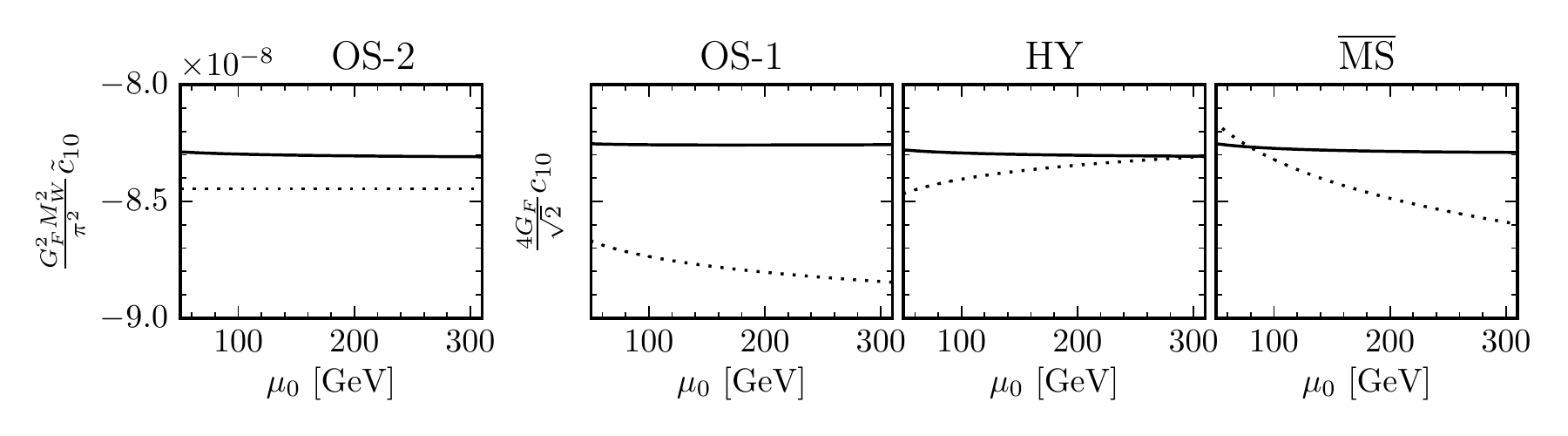}
\end{center}
\caption{Comparison of the matching scale, $\mu_0$, dependence of ${\cal C}_{10}$
  at the scale $\mu_0$ in four renormalization schemes (OS-2, OS-1,
  HY and $\MS$) at LO (dotted) and with NLO~EW corrections (solid). See
  text for more details.}
\label{fig:high-scale:noQCD}
\end{figure*}

We determine the running top-quark mass in the $\overline{\rm MS}$ scheme with
respect to QCD from $M_t^{\rm pole}$ with the aid of the three-loop
relation\footnote{The choice of the matching scale that determines the $N_f = 5$
  to $N_f = 6$ threshold has a numerically negligible impact for $\mu_0 \in [50,
  300]$ GeV considered here.}, $m_t(m_t) = 163.5$~GeV, and evolve it to the
matching scale applying the four-loop expression of the quark-mass anomalous
dimension. Here $m_t$ denotes the top-quark mass, where QCD corrections are
$\MS$-renormalized, but EW corrections are considered in the on-shell scheme. In
the case that the latter are also $\MS$-renormalized, we shall choose the
notation $\overline{m}_t$. The additional shift from $m_t \to \overline{m}_t$,
while numerically quite significant yielding $\overline{m}_t(\overline m_t)
=172.4$~GeV, is dominated by the contribution of tadpole diagrams. The 
tadpole-induced shift cancels in the ratio $x_t =
\overline{m}_t^2/\overline{M}_W^2$
entering the LO Wilson coefficient. 

As already emphasized in \refsec{sec:matching}, once considering higher EW
corrections, the different choices of normalization of the effective Lagrangian
from Eq.~\eqref{eq:c10:norm} affects differently the NLO EW matching corrections of
${\cal C}_{10}$. As renormalization schemes (RS) we consider the on-shell
scheme, the $\MS$ scheme and the hybrid scheme introduced in
\refsec{sec:fullcalc}, which we abbreviate in the following as RS $ = $ OS,
$\MS$ and HY. We apply both, the single-$\GF{}$ and the quadratic-$\GF{}$
normalization for the on-shell scheme denoted as RS $=$ OS-1 and OS-2,
respectively. For RS $ = \MS$ and HY we use only the single-$\GF{}$
normalization.

We first consider the size and the reduction of the scheme dependences in
${\cal C}_{10}$ at the matching scale
\begin{align}
  {\cal C}_{10}(\mu_0) & 
  = \left\{ \begin{array}{l} \displaystyle
     \frac{4 \GF{}}{\sqrt{2}} \tilde{\alpha}_e(\mu_0) \left[
     c_{10}^{(11)} + \tilde{\alpha}_e(\mu_0)\, c_{10}^{(22)}(\mu_0) \right]
  \\[0.5cm]
  \displaystyle
  \frac{\GF{2} M_W^2}{\pi^2} \left[
     \widetilde{c}_{10}^{\,(11)} 
   + \tilde{\alpha}_e(\mu_0)\, \widetilde{c}_{10}^{\,(22)}(\mu_0) \right]
  \end{array} \right. \,,
\end{align}
for the single- and quadratic-$\GF{}$ normalization respectively, after including
the NLO EW corrections ${\cal C}_{10}^{(22)}$. To separate the effects of the EW
calculation, we first switch off any QCD dependence.  Namely, we omit the NLO
QCD correction~${\cal C}_{10}^{(21)}$ and neglect the $\mu_0$ dependence of the
top-quark mass under QCD by fixing the QCD scale and using $m_t(m_t)$ as the
on-shell top-quark mass under EW renormalization, as far as OS-1, OS-2 and HY
schemes are concerned.  In the $\MS$ scheme we perform the additional shift $m_t
\to \overline{m}_t$ using the value of $m_t(m_t)$ as input value. Note, that for
the choice of scale of $m_t$ in the running QCD top mass, the omitted NLO QCD
correction ${\cal C}_{10}^{(21)}$ is particularly small \cite{Buchalla:1993bv,
  Misiak:1999yg, Buchalla:1998ba}, i.e.~the LO result ${\cal C}_{10}^{(11)}$
accounts for the dominant part of the higher-order QCD correction.

The LO and (LO + NLO EW) results are depicted in \reffig{fig:high-scale:noQCD}
for the four renormalization schemes. For $\mu_0$-independent top-quark mass the
LO ${\cal C}_{10}$ is $\mu_0$ independent in the OS-2 scheme, whereas the
replacement $\GF{} \to\alpha_e(\mu_0)/(s_W^{\rm on-shell})^2$ introduces a
$\mu_0$ dependence in OS-1 and a quite significant shift of about 4\% with
respect to OS-2, which translates into a 8\% change of the LO branching ratio.
Although based on the same single-$\GF{}$ normalization, the $\MS$ and HY
schemes exhibit relatively large shifts with respect to OS-1 and a modified
$\mu_0$ dependence due to the $\MS$ renormalization of $s_W$ in both, HY and
$\MS$, schemes and additionally the EW $\MS$ renormalization of the top-quark
and $W$ mass in the $\MS$ scheme.  The overall uncertainty due to EW corrections
at LO may be estimated from the variation of ${\cal C}_{10}$ given by all four
schemes ranging in the interval ${\cal C}_{10}(\mu_0) \in [-8.9,\, -8.2]\cdot
10^{-8}$ for $\mu_0 \in [50, 300]$ GeV corresponding to a $\pm$8\% uncertainty
on the level of the branching ratio. The inclusion of the NLO EW corrections
eliminates this large uncertainty, as all four schemes yield aligned (LO + NLO EW)
results and the $\mu_0$ dependence cancels to large extent in all schemes. The
residual uncertainty due to EW corrections is now confined to the small interval
of ${\cal C}_{10}(\mu_0) \in [-8.31,\, -8.25] \cdot 10^{-8}$ at the scale $\mu_0$,
it is less than $\pm 0.4$\% corresponding to $\pm 0.8$\% on the branching ratio.
The strong reduction of the $\mu_0$ dependence in \reffig{fig:high-scale:noQCD}
is due to the inclusion of NLO corrections in the relation of EW parameters,
which are formally not part of the effective theory and hence cannot be
cancelled by the RGE in the effective theory. At LO in the effective
theory there is no renormalization group mixing of ${\cal C}_{10}$ and the
$\mu_0$ dependence may be used directly as an uncertainty. As discussed in 
\refsec{sec:RGE}, beyond LO in QED the operator mixing will reduce the remaining
$\mu_0$ dependence even further.

Before proceeding, we comment on the OS-1 and $\MS$ scheme and why we shall
discard them for the estimate of residual higher-order uncertainties. The OS-1
scheme exhibits the worst perturbative behavior of all four schemes, as seen in
\reffig{fig:high-scale:noQCD}. The $s_W$-on-shell counterterm induces this, for
an electroweak correction, unnaturally large shift at two-loop. As further
discussed in \refapp{app:OS-1:numerics}, the top-quark mass dependence of the
$s_W$-on-shell counterterm implies a significant higher-order QCD scale
dependence, which we consider artificial. On the other hand, the OS-2 and HY
schemes do not exhibit this strong dependence on the top-quark mass and the
estimate of the size of higher-order QCD contributions by varying the scale of
$m_t$ indicates much smaller corrections. In view of this, we restrict ourselves
to schemes with reasonable convergence properties and leave OS-1 aside. In the
case of the $\MS$ scheme, the application of RG equations is required for the
iterative determination of the EW parameters from the input given in
Eq.~\eqref{eq:EWinput}. For the purpose of \reffig{fig:high-scale:noQCD}, the
presence of QCD could be ignored and lowest-order RG equations were
sufficient. However, in the general case the solution of the according RG
equations are rather involved and we prefer to use the comparison of the HY and OS-2
scheme to estimate higher order EW$\times$QCD corrections.

In the following, we include QCD effects and discuss ${\cal C}_{10}$ at the
low-energy scale $\mu_b$ after applying the RGE running presented in
\refsec{sec:RGE}. We express the Wilson coefficient ${\cal C}_{10}(\mu_b)$ as a
double series in the running couplings $\tilde{\alpha}_s$
and~$\tilde{\alpha}_e$, see Eqs.~\eqref{eq:WC:low:scale} and
\eqref{eq:c10-scaled:exp}, with five relevant contributions ${\cal
  C}_{10,(mn)}$, $(mn = 11,\, 21,\, 02,\, 12,\, 22)$, that depend on Wilson
coefficients of various other operators at the matching scale $\mu_0$. So far,
only the LO $\equiv$ $(mn = 11)$ and the NLO QCD $\equiv$ $(mn = 11 + 21)$
contributions were known. Now, we can include the full NLO EW correction with
the additional contributions $(mn = 11 + 21 + 02 + 12 + 22)$ $\equiv$ NLO (QCD +
EW)\footnote{ These corrections were discussed in the large top-quark-mass limit
  including the RGE effects in Ref.~\cite{Misiak:2011bf}, whereas RGE effects were
  neglected in Ref.~\cite{Buras:2012ru} for  $(mn = 02,12,22)$.}.  For this purpose,
also the scale dependence of $m_t$ that
originates from QCD will be taken into account when varying the matching scale
$\mu_0$.  Note that ${\cal C}_{10}(\mu_b)$ is independent of the matching scale
$\mu_0$ up to the considered orders in couplings due to the inclusion of the RGE
evolution.  However, the residual $\mu_b$ dependence will only be cancelled by
the according $\mu_b$ dependence of the matrix elements of the relevant
operators.

\begin{figure*}[tbp]
\begin{center}
\includegraphics[]{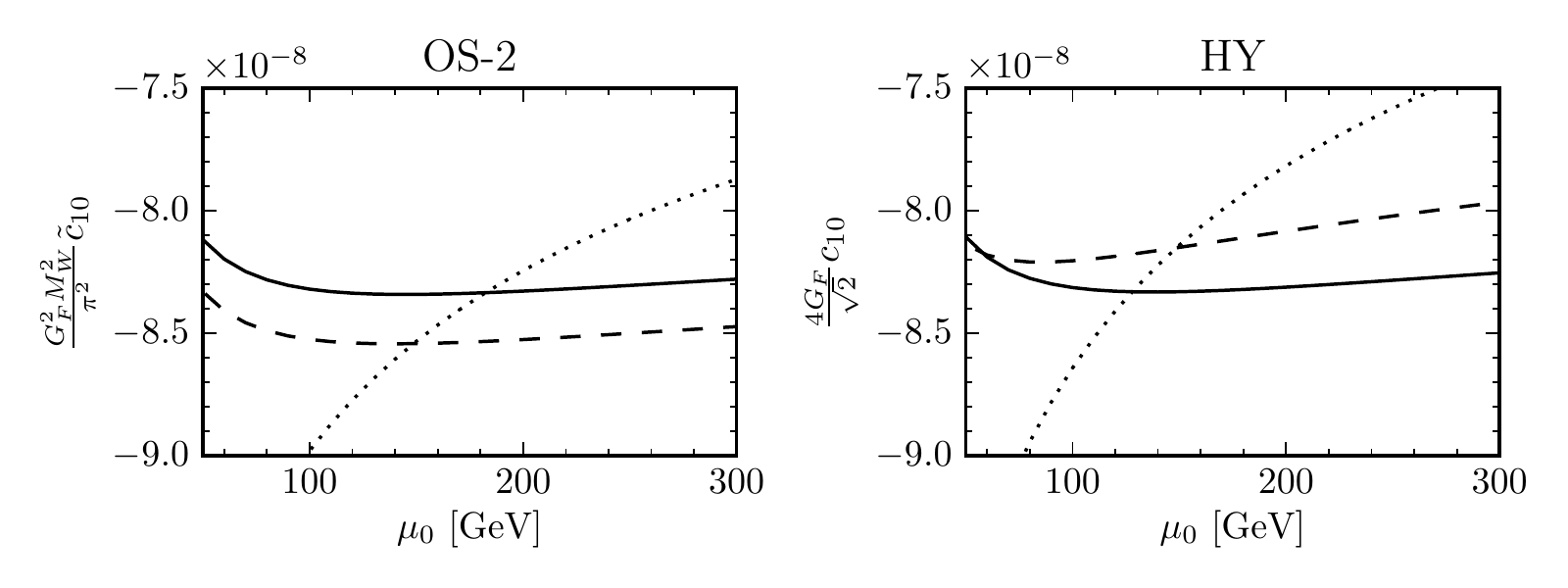}
\end{center}
\caption{The $\mu_0$ dependence of the Wilson coefficient ${\cal C}_{10}
  (\mu_b = 5\, \mbox{GeV})$ in two renormalization schemes (OS-2, HY)
  at LO (dotted), NLO QCD (dashed) and NLO (QCD + EW) (solid). See text
  for more details.}
\label{fig:c10-low:mu0}
\end{figure*}

\begin{figure}[tbp]
\begin{center}
\includegraphics[]{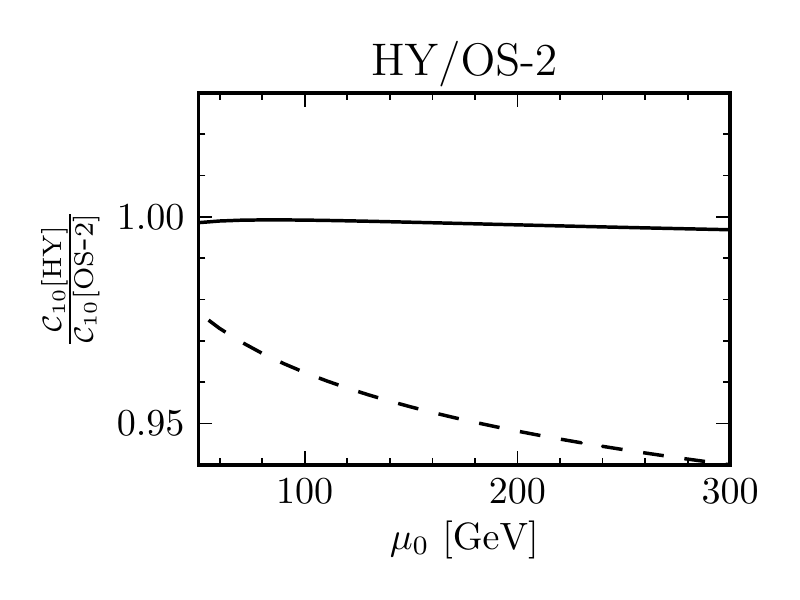}
\end{center}
\caption{The $\mu_0$ dependence of the ratio of the Wilson coefficient ${\cal C}_{10}
  (\mu_b = 5\, \mbox{GeV})$ in the HY and the OS-2 scheme
  at LO and NLO QCD (dashed) and NLO (QCD + EW) (solid). 
  LO and NLO QCD curves coincide.}
\label{fig:c10-low:mu0HYOS2ratio}
\end{figure}

\begin{figure*}[tbp]
\begin{center}
\includegraphics[]{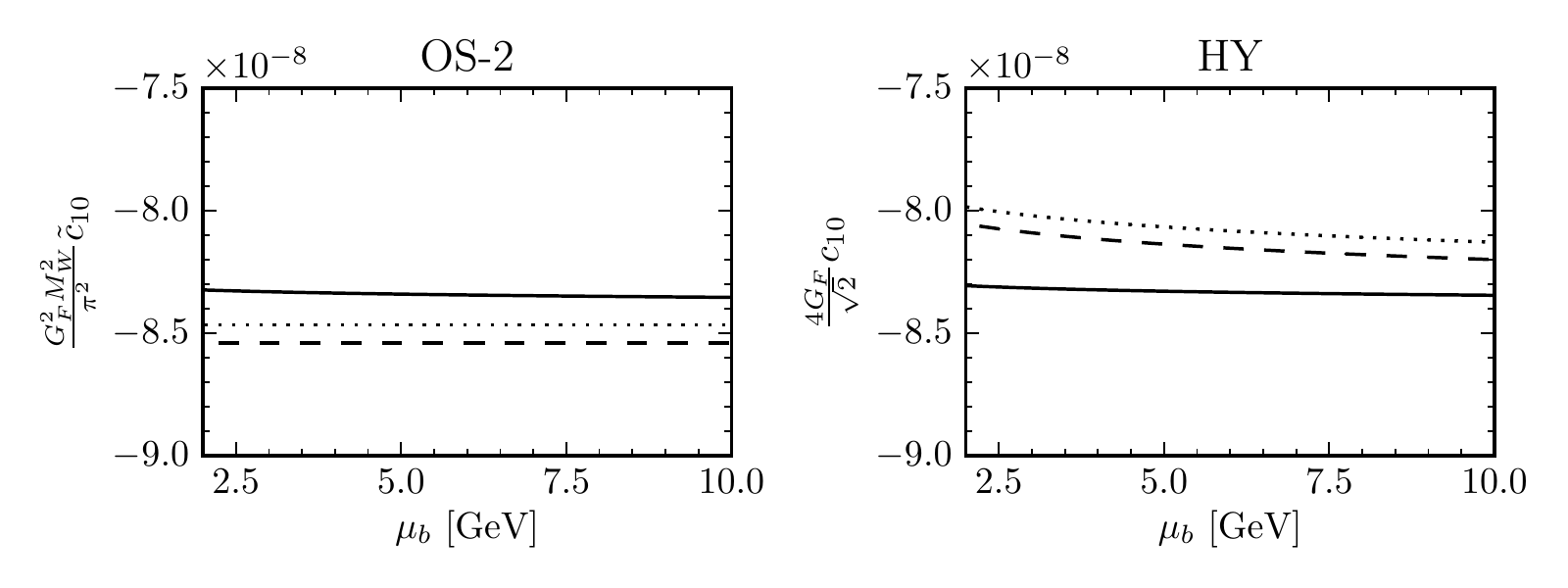}
\end{center}
\caption{The $\mu_b$ dependence of the Wilson coefficient 
  ${\cal C}_{10}(\mu_b)$ for fixed $\mu_0 = 160$ GeV in
  two renormalization schemes (OS-2, HY) at LO (dotted), 
  NLO QCD (dashed) and NLO (QCD + EW) (solid). See text
  for more details.}
\label{fig:c10-low:mub}
\end{figure*}

\reffig{fig:c10-low:mu0} shows the $\mu_0$ dependence of ${\cal C}_{10}(\mu_b =
5\, \mbox{GeV})$ at LO, NLO QCD and NLO (QCD + EW) in the OS-2 and HY
schemes. It is clearly visible that the dependence on the renormalization scale
of $m_t$ reduces when going from LO to NLO QCD and that the LO results coincide
with the ones at NLO QCD at the scale $\mu_0 \approx 150$~GeV.  A further
reduction of this scheme dependence requires the inclusion of NNLO QCD
corrections \cite{Hermann:2013kca}. The NLO QCD result is quite different in the
OS-2 and HY scheme comprising values of ${\cal C}_{10}(\mu_b) \in [-8.54,\,
-7.97] \cdot 10^{-8}$.  The NLO (QCD + EW) result shows again rather large
shifts with respect to NLO QCD and a clear convergence of both schemes towards
the same value. The results of the OS-2 and HY schemes are now confined within
${\cal C}_{10}(\mu_b) \in [-8.34,\, -8.11] \cdot 10^{-8}$ reducing the combined
uncertainty due to scheme dependencies of both QCD and EW interactions to
$\pm$1.4\%. Again, we would like to remind that a substantial part of this
uncertainty is due to so far unknown NNLO QCD corrections. We estimate the
uncertainty due to higher-order EW and QCD corrections to our two-loop EW result
from 1)~the ratio of the results of the HY to the OS-2 scheme, thereby
eliminating the numerically leading QCD $\mu_0$-dependence of $m_t$, and 2)~by
varying the scale $\mu_0$ only in $m_t$ of the two-loop EW matching corrections
$c_{10}^{(22)}$ (or $\widetilde{c}_{10}^{\,(22)}$). As can be seen in
\reffig{fig:c10-low:mu0HYOS2ratio}, at the level of NLO QCD the ratio deviates
quite strongly from $1$ whereas at NLO (QCD + EW) the deviations are less than
$0.3$\%. The ratio of the LO results coincides with the
ratio of the NLO QCD one. We find a similar $\mu_0$ dependence of the OS-2 and
HY results (about $\pm$0.1\%) when varying the scale only in $m_t$ of the EW
two-loop matching correction. We choose the OS-2 scheme with $\mu_0 = 160$~GeV
to predict the central value of ${\cal C}_{10} = -8.341\cdot10^{-8}$, the HY
scheme yields $-8.329\cdot10^{-8}$, and we assign an error due to higher-order EW
corrections from the variation of $\mu_0$ of about $\pm 0.3$\% as suggested by
the comparison of the OS-2 and HY schemes.

We now turn to the discussion of the residual $\mu_b$ dependence for the fixed
value $\mu_0 = 160$~GeV. As already mentioned above, including the according
matrix elements of the involved operators shall decrease this dependence
further, however, for the moment it remains an additional source of uncertainty.
\reffig{fig:c10-low:mub} shows ${\cal C}_{10}(\mu_b)$ at LO, NLO QCD and NLO
(QCD + EW) in the OS-2 and HY schemes. Whereas the values of ${\cal
  C}_{10}(\mu_b)$ are quite different in all three schemes at NLO QCD, the
inclusion of NLO (QCD + EW) corrections in the form of the renormalization group
evolution yields a convergence towards the same value and a very small residual
$\mu_b$ dependence in each scheme of less than $\pm$0.2\%
(OS-2: $\pm$0.16\% and HY: $ \pm$0.20\%)
when varying $\mu_b\in [2.5, 10]$ GeV.  We would like to note, that the
non-perturbative uncertainty due to unknown QED corrections in the evaluation of
the matrix elements is an additional source of uncertainty, not included in the
above estimate.

The dependence of the EW corrections on the Higgs mass is entirely
negligible. Varying $M_H \in [120,\, 130]$~GeV induces variations in ${\cal
  C}_{10}$ of less than $\pm 0.01$\%.

As our final result we choose for the central value the OS-2 scheme with scale
settings $\mu_0 = 160$~GeV and $\mu_b = 5$~GeV
\begin{align}
  \label{eq:final-C10}
  {\cal C}_{10} &= (-8.34\, \pm\, 0.04) \cdot 10^{-8}\,,
\end{align}
where we have estimated higher-order corrections of EW origin from the scale
variations of $\mu_0 \in [50,\, 300]$ GeV and $\mu_b \in [2.5, 10]$ GeV in two
schemes, OS-2 and HY, and added linearly the two errors.  We have not included
into the error budget the residual errors associated to higher QCD corrections
that can be removed by means of the NNLO QCD calculation \cite{Hermann:2013kca} nor
any of the parametric errors listed in \reftab{tab:numinput}. To show
the improvements of our final result \eqref{eq:final-C10}, we quote for
comparison the results at NLO QCD
\begin{align}
  {\cal C}_{10}^{\rm OS-2} & = -8.54 \cdot 10^{-8}\,, &
  {\cal C}_{10}^{\rm HY}   & = -8.14 \cdot 10^{-8}\,  
\end{align}
taken from the according curves of the OS-2 and HY schemes in
\reffig{fig:c10-low:mu0}.

Finally, we compare our prediction with the previous estimate
\cite{Buras:2012ru}, which was obtained using the large-$m_t$ approximation
of ${\cal C}_{10}^{(22)}$ and neglecting the effects of the RGE evolution.  In
particular, the authors found in the HY scheme BR$^{[t=0]} = 3.28 \cdot 10^{-9}$
in Table 2 of their work. Adopting the same numerical input ($f_{B_s} =
227$~MeV, $\tau_{B_s} = 1.466$~ps$^{-1}$, $M_{B_s} = 5.36677$~GeV, $|V_{tb}^{}
V_{ts}^*| = 0.0405$, $m_\mu = 105.6584$~MeV $\Rightarrow$ ${\cal N} = 4.48409
\cdot 10^5$) and Eq.~\eqref{eq:final-C10}, our result BR$^{[t=0]} = 3.13 \cdot
10^{-9}$ is about~$5$\% lower, mainly due to the above mentioned approximations.
Furthermore, the authors of Ref.~\cite{Buras:2012ru} argued that NLO EW
corrections in the HY scheme should be small and suggested a procedure, based on
LO expressions, that lead to the preliminary value of BR$^{[t=0]} = 3.23 \cdot
10^{-9}$ (see Eq.~(17) in Ref.~\cite{Buras:2012ru}), which is closer to our result
and deviates only by~$3$\%. In particular it was suggested to use EW parameters
$\alpha_e$ and $s_W$ in the $\MS$ scheme at the scale $M_Z \approx 90$ GeV and
the LO expression $c_{10}^{(11)} \sim Y_0(x_t)$ with $m_t(m_t)$ with an
additional correction factor $\eta_Y$ to account for higher-order QCD
corrections from $c_{10}^{(21)}$.  We find from \reffig{fig:high-scale:noQCD},
3rd panel for the HY scheme, at $\mu_0 = 90$~GeV a deviation of about 1.5\%
between the LO result and the NLO EW one. We would like to close this comparison
with the remark that the authors of Ref.~\cite{Buras:2012ru} work at LO in the EW
couplings allowing them to combine values of the input parameters which are
dependent beyond the LO, where as in our case certain EW parameters, especially
$M_W$ and $s_W$, do depend on the input quantities of our choice in
Eq.~\eqref{eq:EWinput}. As a consequence, a straightforward numerical comparison
is not possible, however, adopting the suggested procedure using our numerical
values of dependent quantities we obtain a slightly larger value BR$^{[t=0]} =
3.24 \cdot 10^{-9}$ instead of $3.23 \cdot 10^{-9}$. For definiteness we give
here our value $M_W^{\rm on-shell} = (80.358 \pm 0.008)$ GeV obtained with 
\cite{Awramik:2003rn} and our input values, which is close to the current measurement 
$M_W^{\rm PDG} = (80.385 \pm 0.015) $ GeV \cite{Beringer:1900zz}. The largest
uncertainty is due to the variation of $M_t^{\rm pole}$ by $\pm 0.9$~GeV.
Moreover, we use the non-decoupling version for the $\MS$ renormalization
of $s_W^2$ and obtain $s_W^2(M_Z) = 0.2317$ compared to the value $0.2314$
compiled by the PDG \cite{Beringer:1900zz}.

%
%
\section{Conclusions}

We have calculated the next-to-leading (NLO) electroweak (EW) corrections to the
Wilson coefficient ${\cal C}_{10}$ that governs the rare decays $B_q \to
\ell^+\ell^-$ in the Standard Model. To assess the size of higher-order
corrections, the numerical analysis has been performed within three different
renormalization schemes of the involved EW parameters, described in
\refsec{sec:fullcalc}, and two different normalizations of the effective
Lagrangian, given in Eq.~\eqref{eq:c10:norm}. The inclusion of NLO EW
corrections strongly reduced the scheme dependences present at LO for all
considered schemes. We identified the two schemes with the better convergence
behavior and estimated the uncertainty from missing beyond NLO EW corrections to
be about $\pm$0.3\% for ${\cal C}_{10}$. The first renormalization scheme is
based on a new normalization \cite{Misiak:2011bf} that eliminates the ratio
$\alpha_e/s_W^2 \to \GF{}$ in favor of Fermi's constant. The second is based on
the $\MS$ scheme for both quantities entering the ratio $\alpha_e/s_W^2$
\cite{Brod:2010hi}.

Apart from the NLO EW matching corrections to ${\cal C}_{10}$, we took into
account the effects of the renormalization group running of ${\cal C}_{10}$
caused by operator mixing at higher order in QED in the effective theory. As we
do not include QED corrections to the matrix elements of the relevant operators
we estimated the remaining perturbative uncertainty due to the variation of the
low-energy scale $\mu_b$ and found an about $\pm 0.2$\% uncertainty for ${\cal
  C}_{10}$.

In the error budget, we do not include uncertainties due to higher-order QCD
corrections, which are removed by the NNLO QCD calculation \cite{Hermann:2013kca},
nor parametric uncertainties of ${\cal C}_{10}$ and the branching ratio, which
are discussed in detail in Ref.~\cite{Bobeth:2013uxa}.

Our calculation removes an uncertainty of about $\pm 7$\% at the level of the
branching ratio and gives smaller values compared to the conjecture given in
\cite{Buras:2012ru} by about $(3-4)$\%. We have estimated the final
uncertainties due to beyond NLO EW corrections at the matching scale $\mu_0$ and
low-energy scale $\mu_b$. The combination of both results in uncertainties of
$\pm 0.5$\% at the level of ${\cal C}_{10}$ and consequently $\pm 1$\% on the
branching ratio.

\acknowledgments

We would like to thank Joachim~Brod and Andrzej~J.~Buras for many valuable
explanations and suggestions, Bernd~Kniehl for useful correspondence and
Thomas~Hermann, Matthias~Steinhauser and Mikolaj~Misiak for extensive
discussions and careful reading of the manuscript.
Martin Gorbahn acknowledges partial support by the UK Science \& Technology
Facilities Council (STFC) under grant number ST/G00062X/1. Christoph Bobeth
received partial support from the ERC Advanced Grant project ``FLAVOUR'' (267104).

%
%
\appendix

%
%
\section{Details on the Matching Calculation \label{app:technicaldetails}}

%
\subsection{Operator Basis\label{app:operatorbasis}}

Throughout, we use the same definition of the operators as in
Ref.~\cite{Huber:2005ig}.  The RGE evolution from the matching scale $\mu_0$ down to
$\mu_b$ involves the operators mentioned in \refsec{sec:RGE}, whereas here, we
list only operators whose Wilson coefficients contribute to the matching of the
NLO EW correction to ${\cal C}_{10}$ in \refsec{sec:matching}. They are the
physical operator $P_2$ and the according evanescent operator $E_2$
\footnote{Actually, $E_2$ does not contribute to the matching, but only because
  it does not mix in $P_{10}$ at one-loop, i.e.  $\hat Z_{E_2,10}^{(1)}=0$.}
that mediate $b\to q\,\bar{c}c$
\begin{align}
  \label{eq:operatorbasis}
  P_{ 2} & = (\bar q_L \gamma_\mu c_L)\,(\bar c_L \gamma^\mu b_L) \,,
\\[0.1cm]
  E_{ 2} & = (\bar q_L \gamma_{\mu\nu\rho} c_L)\,(\bar c_L \gamma^{\mu\nu\rho} b_L) \,, 
\intertext{as well as $P_9$, $P_{10}$ and the according evanescent operators
 $E_9$ and $E_{10}$ \cite{Bobeth:2003at} that mediate $b\to q\,\ell^+\ell^-$}
  P_{ 9} & = (\bar q_L \gamma_\mu b_L)\sum_\ell (\bar\ell \gamma^\mu \ell) \,,
\\
  P_{10} & = (\bar q_L \gamma_\mu b_L)\sum_\ell (\bar\ell \gamma^\mu\gamma_5 \ell) \,,
\\
  E_{ 9} & = (\bar q_L \gamma_{\mu\nu\rho} b_L)
             \sum_\ell (\bar\ell \gamma^{\mu\nu\rho} \ell) 
             - 10 P_9 + 6 P_{10} \,,
\\
  E_{10} & = (\bar q_L \gamma_{\mu\nu\rho} b_L)
             \sum_\ell (\bar\ell \gamma^{\mu\nu\rho} \gamma_5\ell)  
             + 6 P_9 - 10 P_{10} \,.
\end{align}
The evanescent operators vanish algebraically in $d = 4$ dimensions.
Above $\gamma_{\mu\nu\rho} \equiv \gamma_\mu \gamma_\nu \gamma_\rho$ and 
$\gamma^{\mu\nu\rho} \equiv \gamma^\mu \gamma^\nu \gamma^\rho$. 
In our case, there are no equation-of-motion vanishing operators with a
projection on $\langle P_{10}\rangle^{(0)}$ to contribute to the matching.

%
\subsection{Details on the Standard Model Calculation \label{app:fullcalc}}

The two-loop EW SM calculation is very similar to the analogous calculation for
the $K\to\pi\nu\bar{\nu}$ decays \cite{Brod:2010hi}. The calculation comprises
of generating and calculating all two-loop topologies for the transition $b\to
q\ell^+\ell^-$ (Fig.~\ref{fig:SMdiagrams}).

We perform two independent calculations, in the first we use {\tt FeynArts}
\cite{Hahn:2000kx} to generate the topologies and a self-written {\tt
  Mathematica} program to evaluate them and in the second {\tt QGRAF}
\cite{Nogueira:1991ex} and a self-written {\tt FORM}
\cite{Kuipers:2012rf} program, respectively.

By setting the external momenta and the masses of all fermions except for the
top quark to zero all diagrams reduce to massive tadpoles with maximally three
different masses. We reduce them to a few known master integrals using the
recursion relations from Refs.~\cite{Bobeth:1999mk,Davydychev:1992mt}.

We work in dimensional regularization, which raises the question of
how to treat $\gamma_5$ in $d \neq 4$ dimensions. The naive
anticommutation relation (NDR) $\{\gamma_5,\gamma_\mu\}=0$ can lead to
algebraic inconsistencies in the evaluation of traces with
$\gamma_5$'s. Yet, the algebraically consistent definition of
$\gamma_5$ by 't\,Hooft-Veltman (HV) \cite{'tHooft:1972fi} leads to
spurious breaking of the axial-current Ward identities and as such
requires the incorporation of symmetry-restoring finite
counterterms. Diagrams that are free of algebraic inconsistencies in
the NDR scheme yield the same finite result after the appropriate
counterterms are added. This trivially holds 
for all diagrams free of internal fermion loops as well as for diagrams 
that involve traces with an even number of $\gamma_5$ matrices if 
the $\gamma_5$ matrices are eliminated through naive anticommutation 
from the relevant traces \cite{Larin:1993tq}. 
Since selfenergy diagrams involving a single 
axial coupling vanish, diagrams involving fermionic loops on
bosonic propagators also correspond to the same finite expression in
both schemes after appropriate renormalisation. 
Accordingly, special care has to be taken only for diagrams 
involving a fermion-triangle loop and coming with an odd number 
of $\gamma_5$ matrices. 
We use the HV prescription for these type of diagrams, since in particular
the diagram with three $\gamma_5$ matrices cannot be simply calculated
in the NDR scheme. 
Here we note that the finite renormalization, which will restore 
the axial-anomaly relation of diagrams involving fermion traces,
will drop out in our calculation after the sum over the complete 
set of standard model fermions is performed. 
This follows from the fact that Standard Model is anomaly free and can 
be understood by noting that e.g.\ the difference of the
singlet and non-singlet counterterm in Ref.~\cite{Larin:1993tq} has
opposite sign for up-type and down-type quarks. 
Yet, one subtlety could arise from charged $W$ and Goldstone bosons 
connecting the fermion-triangle diagram with the external fermion line. 
The axial couplings on the external line could in principle result in 
a spurious breaking of the axial-current Ward identity if treated in 
the HV scheme. 
Yet, only the 4-dimensional part of this coupling contributes 
if the fermion triangle contains an odd number of $\gamma_5$ matrices, since the
corresponding diagrams are either finite after GIM or their traces
vanish. 
Accordingly, we can safely use the HV scheme in these
circumstances without the need of an extra finite renormalisation and
calculate all other diagrams in the NDR scheme.
The effective theory 
calculation does not involve fermion traces with $\gamma_5$ and for this reason
can be performed completely in the NDR scheme.

In the SM, the renormalization scheme of the fermion fields $f = q,\, \ell$,
i.e.\ quarks and leptons, is chosen such that the kinetic terms in the effective
theory remain canonically normalized at NLO in EW interactions. As a consequence,
Wilson coefficients of dimension three $b\to s$ mediating operators in the effective
theory are zero. The bare SM fields, $f^{(0)}$, with flavor type $i$ and of 
chirality-type $a$ are renormalized 
\begin{align}
  f^{(0)}_{i,a} = \left( \delta_{ij} + \frac{1}{2} Z_{ij}^a\right) f_{j,a}
\end{align}
with the help of the matrix-valued field renormalization constant $Z^a$. The
latter is determined from one-loop $f \to f'$ two-point functions such that the
matching relation for the fields in the SM and effective theory
\begin{align}
  f^{\rm full} & = f^{\rm eff} \,,
\end{align}
holds, implying that tree-level matrix elements of operators, $\langle P_i
\rangle^{(0)}$, are the same in the SM and effective theory amplitude, see
Eqs.~\eqref{eq:afull} and \eqref{eq:eff-th:amp} respectively. For this purpose,
the two-point functions are evaluated in an expansion up to first order in
external momenta and masses over heavy masses. The heavy particle contributions
yield finite parts to $Z^a$, whereas light particle contributions eventually drop
out in the matching and thus may be discarded in the calculation. In addition,
the flavor off-diagonal quark-field renormalization constant $Z_{bq}$ is determined
at two-loop level from the two-point function $b\to q$. 

The counterterm of the CKM matrix is entirely determined by the field
renormalization constants $Z^L$ of the up- and down-quark fields.
This renormalization prescription corresponds to a definition of the
CKM elements in the effective theory where the kinetic terms of all
light quark fields are canonical.

Since we renormalize both the couplings $\alpha_e^{\rm full}$ and $\alpha_e^{\rm
  eff}$ of the full and effective theory, respectively, in the $\MS$ scheme, the
$\alpha_e$ threshold corrections have to be included in the case of the
single-$\GF{}$ normalization. In the threshold corrections, $\Delta \alpha_e$,
\begin{equation}
  \label{eq:alpha_e:threshold}
\begin{aligned}
  \alpha_e^{\rm full} & 
  = \alpha_e^{\rm eff} \, \left[1 + \frac{\alpha_e^{\rm eff}}{4\pi} \Delta \alpha_e \right]\,,
\\[0.2cm]
  \Delta \alpha_e &
  = - \frac{2}{3} - 14 \ln \frac{\mu}{M_W} + \frac{32}{9} \ln \frac{\mu}{M_t}
\end{aligned}
\end{equation}
the first two terms arise from the decoupling of the electroweak gauge bosons 
and the last term from the top quark at the scale $\mu$. Since the definition
of  $\alpha_e(M_Z)$ in \reftab{tab:numinput} compiled by the particle data group
\cite{Beringer:1900zz} already implies a decoupled top quark, we determine 
$\alpha_e^{\rm eff}$ from $\alpha_e(M_Z)$ using only the gauge boson contribution
and find $\alpha_e^{\rm eff}(M_Z) = 1/127.751$ that we use in our numerical
evaluations.

In order to match consistently, we apply
Eq.~\eqref{eq:alpha_e:threshold} to substitute the $\alpha_e^{\rm full} \to
\alpha_e^{\rm eff}$, which affects the matching at next-to-leading order due to
an additional contribution in the amplitude of the full theory from the lower
order part in Eq.~\eqref{eq:afull:series} (omitting here the subscript $A_{{\rm
    full},10} \to A$)
\begin{equation}
\begin{aligned}
  & \tilde{\alpha}_e^{\rm full} A^{(1)}  
  + \left(\tilde{\alpha}_e^{\rm full}\right)^2 A^{(2)} =
\\[0.2cm] 
  & 
    \tilde{\alpha}_e^{\rm eff} A^{(1)} + 
   \left(\tilde{\alpha}_e^{\rm eff}\right)^2 \left[ 
     A^{(2)} +
     \Delta \alpha_e \, A^{(1)} \right]\,.
\end{aligned}
\end{equation}

%
\subsection{Details on the Effective Theory Calculation \label{app:effcalc}}

Before being able to evaluate the two-loop $b\to q\ell^+\ell^-$ amplitude in the
effective theory we need to know all Wilson coefficients and renormalization
constants appearing in Eq.~\eqref{eq:A2eff}. The tree-level contribution ${\cal
  C}_{2}^{(00)}$ and the one-loop results ${\cal C}_{9}^{(11)}$ and ${\cal
  C}_{10}^{(11)}$ are given in Ref.~\cite{Bobeth:1999mk} including the ${\cal
  O}(\epsilon)$ terms for the latter two. Here we give in addition the Wilson
coefficients of the two evanescent operators
\begin{equation}
\begin{aligned}
  c_{E_9}^{(11)} & = c_{E_{10}}^{(11)} =
\\[0.2cm]
  & \frac{1}{16 s_W^2} \frac{x_t}{(x_t-1)^2} \left( 1 - x_t + \log x_t \right)
   + {\cal O}(\epsilon) \,.
\end{aligned}
\end{equation}  
The ${\cal O}(\epsilon)$ terms of $c_{E_9}^{(11)}$ and $c_{E_{10}}^{(11)}$ do
not contribute to the matching\footnote{The operator $E_{10}$ does not contribute
to the matching at all because $\hat Z_{E_{10}, 10}^{(1)} = 0$.} as the mixing
renormalization constants $\hat Z_{E_{9}, {10}}^{(1)}$ and $\hat Z_{E_{10}, {10}}^{(1)}$
carry no divergent terms, only finite ones.

Having all relevant Wilson coefficients we return to the renormalization
constants. We fix the field renormalization constants by extracting the
UV poles of the appropriate photonic one-loop two-point functions in the five-flavor
theory. The results are:
\begin{equation}
  \label{eq:eff-th:ren-constants}
  Z_{i}^{} 
  = 1 + \tilde\alpha_e \, Z_{i}^{(1)} + \dots 
\end{equation}
with 
\begin{align*}
  Z_{d}^{(1)} & = -\frac{1}{9\epsilon} \,, & 
  Z_{\ell}^{(1)} & = -\frac{1}{\epsilon}\,.
\end{align*}
We proceed similarly for the constants governing the mixing of operators into
$P_{10}$. We calculate the UV poles of all one-loop insertions of a given
operator, project on the tree-level matrix element of $P_{10}$ and absorb the
left-over pole in the mixing renormalization constant.

For the case of physical operators mixing into physical ones we absorb only the
divergences into the constants $\hat Z_{P,P}$. For evanescent operators this is
not the case. Evanescent operators are unphysical in four dimensions and at each
order in perturbation theory their operator basis needs to be extended. To
ensure that the Wilson coefficients at a given fixed order are independent from
the choice of evanescent operators in some higher order we include finite terms
in $\hat Z_{E,P}$ and completely cancel the mixing of evanescent to physical
operators.

We have calculated all contributing one-loop mixing renormalization
constants including the mixing of evanescent to physical operators. 
The mixing of physical operators can also be extracted from the 
anomalous dimension matrices in Refs.~\cite{Bobeth:2003at, Huber:2005ig}.
Here we report the relevant non-zero constants 
\begin{align}
  \hat Z_{9, 10}^{(1)} & = -\frac{2}{\epsilon} \,, &
  \hat Z_{E_{ 9}, 10}^{(1)} & = \frac{32}{3} \,.
\end{align}
We extract the $1/\epsilon$-part of the one two-loop renormalization constant we
need from the corresponding anomalous dimension in Ref.~\cite{Bobeth:2003at} and
calculated the $1/\epsilon^2$-term
\begin{equation}
  \label{eq:Z2:210}
  \hat Z_{2, 10}^{(2)} 
  = \frac{4}{9\epsilon^2} -\frac{26}{27\epsilon} \,.
\end{equation}

%
%
\section{Details on the RGE\label{app:RGE}}

%
\subsection{General}

The dependence of the Wilson coefficients ${\cal C}_i$ on the renormalization
scale $\mu$ is governed by the anomalous dimension matrix $\hat{\gamma}$
\begin{align}
  \mu \frac{d}{d\mu} {\cal C}_i(\mu) & 
  = \left[\hat{\gamma}^T (\mu)\right]_{ij} {\cal C}_j(\mu)
\end{align}
with the expansion in the couplings
\begin{align}
  \hat{\gamma} (\mu) = 
    \sum_{\substack{m,n = 0 \\ m+n \geq 1}} 
    \tilde{\alpha}_s(\mu)^m \tilde{\alpha}_e(\mu)^n\, \hat{\gamma}_{(mn)} \,,
\end{align}
which is known up to and including relevant entries in $(mn) = (30)$ and $(21)$.
It has been solved as an expansion in terms of the small quantities
\cite{Huber:2005ig}
\begin{align}
  \omega  & \equiv 2 \beta^s_{00}\, \tilde{\alpha}_s(\mu_0),
\\[0.2cm]
  \lambda & \equiv 
    \frac{\beta^e_{00}}{\beta^s_{00}} \frac{\tilde{\alpha}_e(\mu_0)}{\tilde{\alpha}_s(\mu_0)}
    = \frac{\beta^e_{00}}{\beta^s_{00}} \kappa(\mu_0)
\end{align}
in which case the evolution operator in Eq.~\eqref{eq:RGE:general:sol} takes the form
\begin{align}
  \label{eq:U:lam-om:exp}
  U(\mu_b, \mu_0) & = \sum_{m, n \geq 0}^{2} \omega^m \lambda^n \, U_{(mn)} \,,
\end{align}
excluding the term $(mn) = (22)$ that requires the knowledge of higher-order
contributions to the anomalous dimension matrix. The $U_{(mn)}$ can be read off from Eq.~(47) of
Ref.~\cite{Huber:2005ig}, whereas the initial Wilson coefficients (in the single-$\GF{}$
normalization) at the scale $\mu_0$ have the expansion
\begin{equation}
  \label{eq:iWC:lam-om:exp}
\begin{aligned}
  c_i(\mu_0) & = 
    c_i^{(00)} + \omega \frac{c_i^{(10)}}{2 \beta^s_{00}} 
  + \omega^2 \frac{c_i^{(20)}}{(2 \beta^s_{00})^2}
\\[0.2cm] &
  + \omega\lambda \frac{c_i^{(11)}}{2 \beta^e_{00}} 
  + \omega^2\lambda \frac{c_i^{(21)}}{4 \beta^e_{00} \beta^s_{00}}
  + \omega^2\lambda^2 \frac{c_i^{(22)}}{(\beta^e_{00})^2} \,.
\end{aligned}
\end{equation}
The components ${\cal C}_{i,(mn)}$ of the downscaled Wilson coefficients
in Eq.~\eqref{eq:WC:low:scale} are then obtained from the reexpansion of
Eq.~\eqref{eq:RGE:general:sol} in the new parameters $\tilde{\alpha}_s(\mu_b)$
\begin{align}
  \omega & = 2 \beta^s_{00}\, \eta\, \tilde{\alpha}_s(\mu_b)\,, &
\end{align}
and $\kappa(\mu_b)$
\begin{equation}
\begin{aligned}
  \lambda & = \frac{\beta^e_{00}}{\beta^s_{00}} \frac{\kappa(\mu_b)}{\eta}
    \Big[1 + \kappa(\mu_b) A_1(\eta)
\\ &  \hskip 1.5cm +  \tilde{\alpha}_s(\mu_b) \kappa(\mu_b) A_2(\eta)
        + {\cal O}\left(\kappa^2, \,\tilde{\alpha}_s^2\right)\Big]
\end{aligned}
\end{equation}
after inserting Eqs.~\eqref{eq:U:lam-om:exp} and \eqref{eq:iWC:lam-om:exp}. The
coefficients $A_{1,2}(\eta)$ are given in Eq.~(67) of Ref.~\cite{Huber:2005ig}.

%
\subsection{Solution \label{app:RGE:sol}}

\begin{table*}
\begin{center}
\renewcommand{\arraystretch}{1.2}
\begin{tabular}{c|cccccccc}
 $i$ 
& 1 & 2 & 3 & 4 & 5 & 6 & 7 & 8
\\
\hline
$b_i$ 
& $\phantom{+} 0.00354$ 
& $\phantom{+} 0.01223$
& $-0.00977$
& $-0.01070$
& $-0.00572$ 
& $\phantom{+} 0.00022$
& $\phantom{+} 0.01137$
& $-0.00117$
\\[1em]
  $d_i^{(2a)}$
& $0$
& $0$
& $\phantom{+} 0.61602$ 
& $\phantom{+} 0.44627$ 
& $\phantom{+} 0.57472$
& $\phantom{+} 0.08573$
& $-0.48807$ 
& $-0.24089$
\\
  $d_i^{(2b)}$
& $-1.18162$
& $\phantom{+} 0.22940$
& $\phantom{+} 0.06522$ 
& $-0.04380$ 
& $-0.02201$
& $-0.00316$
& $-0.03366$ 
& $-0.00414$
\\[1em]
  $d_i^{(1)}$
& $\phantom{+} 0.01117$
& $-0.03088$
& $\phantom{+} 0.00411$ 
& $\phantom{+} 0.00713$ 
& $\phantom{+} 0.00478$
& $\phantom{+} 0.00012$
& $\phantom{+} 0.00379$ 
& $-0.00023$
\\
  $d_i^{(4)}$
& $-0.00799$
& $-0.03666$
& $\phantom{+} 0.06300$
& $ 0$
& $-0.01519$
& $-0.00071$
& $ 0$
& $-0.00344$
\\[1.5em]
  $e_i^{(1a)}$
& $0$
& $0$
& $-0.25941$ 
& $-0.29751$ 
& $-0.48014$
& $\phantom{+} 0.04647$
& $-0.16269$ 
& $-0.04728$
\\
  $e_i^{(1b)}$
& $\phantom{+} 1.13374$
& $\phantom{+} 0.09381$
& $-0.03041$ 
& $\phantom{+} 0.00781$ 
& $\phantom{+} 0.01838$
& $-0.00138$
& $-0.02259$ 
& $\phantom{+} 0.00121$
\\[1em]
  $e_i^{(4a)}$
& $0$
& $0$
& $-4.03683$ 
& $0$ 
& $\phantom{+} 1.52565$
& $-0.27461$
& $0$ 
& $-0.70642$
\\
  $e_i^{(4b)}$
& $\phantom{+} 3.38669$
& $-0.10885$
& $\phantom{+} 0.16283$ 
& $0$ 
& $\phantom{+} 0.06697$
& $-0.01681$
& $0$ 
& $\phantom{+} 0.00137$
\\[1em]
  $e_i^{(1)}$
& $\phantom{+} 0.01117$
& $-0.03088$
& $\phantom{+} 0.00411$ 
& $\phantom{+} 0.00713$ 
& $\phantom{+} 0.00478$
& $\phantom{+} 0.00012$
& $\phantom{+} 0.00379$ 
& $-0.00023$
\\
  $e_i^{(2)}$
& $\phantom{+} 0.00354$
& $\phantom{+} 0.01223$
& $-0.00977$ 
& $-0.01070$ 
& $-0.00572$
& $\phantom{+} 0.00022$
& $\phantom{+} 0.01137$ 
& $-0.00117$
\\
  $e_i^{(3)}$
& $\phantom{+} 0.02179$
& $-0.12336$
& $\phantom{+} 0.07870$ 
& $0$ 
& $\phantom{+} 0.01930$
& $\phantom{+} 0.00873$
& $0$ 
& $-0.00516$
\\
  $e_i^{(4)}$
& $-0.00799$
& $-0.03666$
& $\phantom{+} 0.06400$ 
& $0$ 
& $-0.01519$
& $-0.00071$
& $0$ 
& $-0.00344$
\\
  $e_i^{(5)}$
& $\phantom{+} 0.19550$
& $-0.93249$
& $\phantom{+} 0.37858$ 
& $0$ 
& $\phantom{+} 0.39909$
& $\phantom{+} 0.05921$
& $0$ 
& $-0.09989$
\\
  $e_i^{(6)}$
& $-0.17154$
& $\phantom{+} 0.39616$
& $\phantom{+} 0.01201$
& $0$ 
& $-0.19423$ 
& $\phantom{+} 0.00357$
& $0$ 
& $-0.04597$
\end{tabular}
\renewcommand{\arraystretch}{1.0}
\end{center}
\caption{Numerical values of $b_i^{}$, $d_i^{(j)}$ and $e_i^{(j)}$ 
  entering \eqref{eq:c10:RGE:solution}.
}
\label{tab:magic:numbers}
\end{table*}

Here the solution of the components $c_{10,(mn)}$ in Eq.~\eqref{eq:RGE:general:sol} of
the single-$\GF{}$ normalization from Eq.~\eqref{eq:DeltaB1:eff:Lag} at the low scale
$\mu_b$ are given in terms of $\eta = \alpha_s(\mu_0)/\alpha_s(\mu_b)$ and their
initial components $c_i^{(mn)}$ in Eq.~\eqref{eq:WC:matching:scale} at the matching
scale $\mu_0$. The derivation of the according results $\widetilde c_{10,(mn)}$
for the quadratic-$\GF{}$ normalization was given in \refsec{sec:RGE}.

The numerical diagonalization of the leading-order anomalous dimension yields
the exponents
\begin{equation}
\begin{aligned}
  a_i & 
  = (-2,\, -1,\, -0.899395,\, -0.521739,
\\ & \quad\,\,\, -0.422989,\, 
     0.145649,\, 0.260870,\, 0.408619)\,.
\end{aligned}
\end{equation}
The components read
\begin{widetext}
\begin{equation}
\begin{split}
\label{eq:c10:RGE:solution}
  c_{10,(11)} & = c_{10}^{(11)},\qquad\qquad\quad
  c_{10,(21)}   = \eta\, c_{10}^{(21)},\qquad\qquad\quad 
  c_{10,(02)}   = \sum_{i=1}^8 b_i \eta^{a_i} \, c_{2}^{(00)},\\[2em] 
  c_{10,(12)} & = 
   		  \sum_{i=1}^8 \eta^{a_i + 1} 
    		  \Big[ \left( d_i^{(2a)} \eta^{-1} + d_i^{(2b)} \right) c_{2}^{(00)}
       		  + d_i^{(1)} c_{1}^{(10)} + d_i^{(4)} c_{4}^{(10)} \Big]\\[0.2cm]
  	      & \quad - 0.11060 \frac{\ln\eta}{\eta} c_{2}^{(00)} 
    		      + \left(\eta^{-1} -1\right)
   		      \left(0.26087\, c_{9}^{(11)} + 1.15942\, c_{10}^{(11)}\right),\\[2em]
  c_{10,(22)} & = 
  		  \sum_{i=1}^8 \eta^{a_i + 2} \left[ 
          	  \left( e_i^{(1a)} \eta^{-1} + e_i^{(1b)} \right) c_{1}^{(10)}
        	  + \left( e_i^{(4a)} \eta^{-1} + e_i^{(4b)} \right) c_{4}^{(10)} 
        	  + \sum_{j=1}^6 e_i^{(j)} c_{j}^{(20)}\right]\\[0.2cm]
  	      & \quad + \left(0.27924\, c_{1}^{(10)} + 0.33157\, c_{4}^{(10)} 
                      + 2.35917\, c_{9}^{(11)} + 3.29679\, c_{10}^{(11)} \right) \ln\eta\\[0.2cm]
  	      & \quad + \left(1 - \eta \right) \left(0.26087\, c_{9}^{(21)} 
    		      + 1.15942\, c_{10}^{(21)}\right)
    		      + c_{10}^{(22)},
\end{split}
\end{equation}
with the coefficients $b_i^{}$, $d_i^{(j)}$ and $e_i^{(j)}$ given in
\reftab{tab:magic:numbers}. 

\end{widetext}

%
%
\section{Numerical study of ${\cal C}_{10}$ in OS-1 scheme
  \label{app:OS-1:numerics}}

\begin{figure}[]
\begin{center}
\includegraphics[]{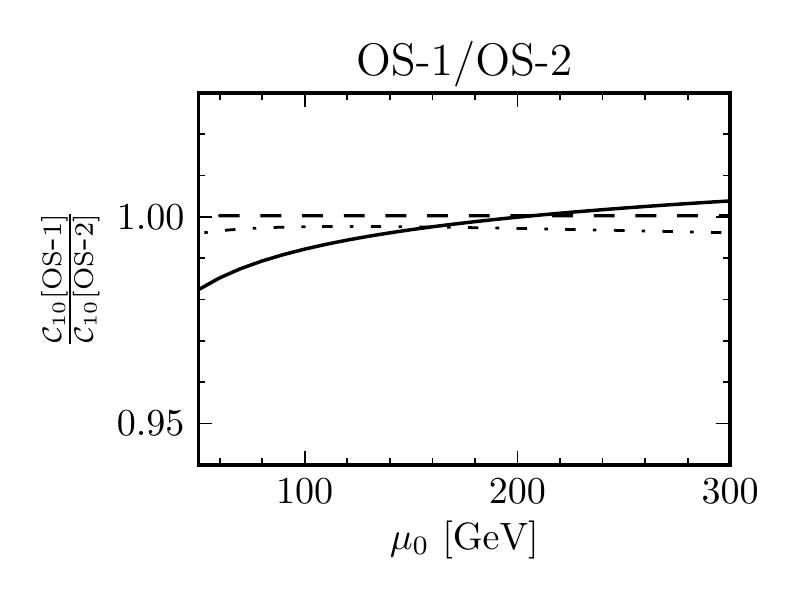}
\end{center}
\caption{The $\mu_0$ dependence of the ratio of the Wilson coefficient ${\cal
    C}_{10} (\mu_b = 5\, \mbox{GeV})$ in OS-1 and OS-2 schemes.  The LO and NLO
  QCD result coincide (dashed). The full $\mu_0$ dependence of NLO (QCD + EW)
  (solid) and partial $\mu_0$ dependence for fixed $m_t(160\, \mbox{GeV})$ in
  the $s_W$-on-shell counterterm (dashed dotted).  }
\label{fig:c10-low:mu0OSOS2ratio}
\end{figure}

In this appendix we estimate higher-order corrections in the OS-1 scheme and
supplement in this context the discussion of the OS-2 and HY schemes from
\refsec{sec:num:analysis}. For this purpose, we proceed as in
\reffig{fig:c10-low:mu0} and \reffig{fig:c10-low:mu0HYOS2ratio} and vary the
matching scale $\mu_0$, which allows to estimate higher-order QCD corrections
via the dependence on the running top-quark mass. The result is shown in
\reffig{fig:c10-low:mu0OSOS2ratio} at NLO QCD and NLO (EW + QCD) order
normalized to the OS-2 result at the respective orders. To understand the
different $\mu_0$ dependence of the NLO QCD result for the OS-1 and OS-2
schemes, we remind that they involve different normalizations (see Eq.~\eqref{eq:C10LO}),
which bear a $\mu_0$ dependence due to their $m_t$ dependence when determining
values of $M_W^{\rm on\textrm{-}shell}$ and consequently $s_W^{\rm
  on\textrm{-}shell}$, see Eq.~\eqref{eq:swonshell} and the input in
Eq.~\eqref{eq:EWinput}. As mentioned in \refsec{sec:fullcalc}, we calculate
$M_W^{\rm on\textrm{-}shell}$ with the aid of the result in Ref.~\cite{Awramik:2003rn},
which incorporates various higher-order corrections that contribute beyond the
NLO EW calculation of ${\cal C}_{10}$ performed in this work, especially those
that require the choice of a particular renormalization scheme for the top-quark
mass. Throughout we use the pole top mass as numerical input as in 
Ref.~\cite{Awramik:2003rn}.

At NLO (EW + QCD) the OS-1 scheme exhibits a very different $\mu_0$ dependence
with respect to OS-2 and HY schemes, which is increased compared to NLO QCD.
The main reason being the large EW two-loop correction to $c_{10}^{(22)}$ from
the $s_W$-on-shell counterterm as already mentioned in connection with
\reffig{fig:high-scale:noQCD}. The counterterm has a strong top-quark-mass
dependence.  To illustrate the latter, we present in
\reffig{fig:c10-low:mu0OSOS2ratio} additionally the NLO (EW + QCD) result
(dashed-dotted line) when keeping the scale of the running top-quark mass in the
counterterm contribution fixed at $\mu_0 = 160$ GeV.  Hence, the large shift
caused by the electroweak two-loop correction in the OS-1 scheme is accompanied
with an artificially large top-quark-mass dependence.  As a consequence we do
not consider the OS-1 scheme in our estimate of higher-order uncertainties. It
would increase the estimate due to $\mu_0$ variation of about $\pm 0.3$\% given
in \refsec{sec:num:analysis} to about $+ 0.4$\% and $-1.7$\%.

%
%
\bibliographystyle{apsrev4-1}
\bibliography{refs}

\end{document}